\documentclass[10pt,twocolumn,letterpaper,notables]{article}
\usepackage{cvpr}
\usepackage{cvpr}      

\usepackage{graphicx}
\usepackage{amsmath}
\usepackage{amssymb}
\usepackage{booktabs}

\usepackage[pagebackref,breaklinks,colorlinks]{hyperref}

\usepackage[capitalize]{cleveref}
\crefname{section}{Sec.}{Secs.}
\Crefname{section}{Section}{Sections}
\Crefname{table}{Table}{Tables}
\crefname{table}{Tab.}{Tabs.}

\def\cvprPaperID{*****} 
\def\confName{CVPR}
\def\confYear{2024}

\usepackage{tikz}
\usepackage{booktabs,lmodern}
\usepackage{overpic}
\usepackage{graphicx}
\usepackage{xcolor, colortbl}
\usepackage{tabularx}
\usepackage[export]{adjustbox}
\usepackage{multirow}
\usepackage{tcolorbox}
\usepackage{arydshln}
\usepackage{mathtools}
\definecolor{c1}{RGB}{128, 237, 18}
\definecolor{c2}{RGB}{146, 225, 11}
\definecolor{c3}{RGB}{165, 214, 4}
\definecolor{c4}{RGB}{182, 198, 2}
\definecolor{c5}{RGB}{199, 182, 1}
\definecolor{c6}{RGB}{213, 164, 5}
\definecolor{c7}{RGB}{227, 146, 9}
\definecolor{c8}{RGB}{236, 127, 18}
\definecolor{c9}{RGB}{246, 108, 28}
\definecolor{c10}{RGB}{236, 127, 18}
\definecolor{tabfirst}{rgb}{1, 0.7, 0.7} 
\definecolor{tabsecond}{rgb}{1, 0.85, 0.7} 
\definecolor{tabthird}{rgb}{1, 1, 0.7} 

\makeatletter
\newif\if@notables
\relax
\newcommand*{\ifnotables}{%
  \if@notables
    \expandafter\@firstoftwo
  \else
    \expandafter\@secondoftwo
  \fi
}
\makeatother

\newcommand{\acronymtitle}{\textcolor{c1}{D}\textcolor{c2}{i}\textcolor{c3}{f}\textcolor{c4}{f}\textcolor{c5}{u}\textcolor{c6}{s}\textcolor{c7}{e}\textcolor{c8}{R}\textcolor{c9}{A}\textcolor{c10}{W}}
\newcommand{\acronym}{DiffuseRAW}
\newcommand{\psnr}{PSNR $\uparrow$}
\newcommand{\ssim}{SSIM $\uparrow$}
\newcommand{\lpips}{LPIPS $\downarrow$}
\usepackage{etoolbox}

\newcolumntype{L}{>{\raggedright\arraybackslash}X}
\newcolumntype{R}{>{\raggedleft\arraybackslash}X}

\usepackage{amsmath,amsfonts,bm}
\usepackage{mathtools}

\def\eqref#1{equation~\ref{#1}}

\def\1{\bm{1}}

\DeclareMathAlphabet{\mathsfit}{\encodingdefault}{\sfdefault}{m}{sl}
\SetMathAlphabet{\mathsfit}{bold}{\encodingdefault}{\sfdefault}{bx}{n}

\newcommand{\E}{\mathbb{E}}

\newcommand{\Eb}[2]{\E_{#1}\!\left[#2\right]}

\newcommand{\bc}{\mathbf{c}}

\newcommand{\bx}{\mathbf{x}}

\newcommand{\bz}{\mathbf{z}}

\newcommand{\bepsilon}{{\boldsymbol{\epsilon}}}

\begin{document}

\title{\acronymtitle: End-to-End Generative RAW Image Processing for Low-Light Images}

\author{
Rishit Dagli\\
University of Toronto\\
{\tt\small rishit@cs.toronto.edu }
}


\setlength{\tabcolsep}{3pt}
\setlength{\fboxrule}{.1pt}
\renewcommand{\arraystretch}{1}
\twocolumn[{%
\renewcommand\twocolumn[1][]{#1}%
\maketitle
\begin{center}
    \centering
    \captionsetup{type=figure}
    \begin{tabular}{ccccc}
        & Camera Output & Scaled RAW Images & Traditional Pipeline & \acronym \ on RAW data \\
         \multirow{2}{*}[12ex]{\rotatebox[origin=c]{90}{$\times250$, ISO320 (outdoor)}} & \frame{\includegraphics[width=0.23\textwidth]{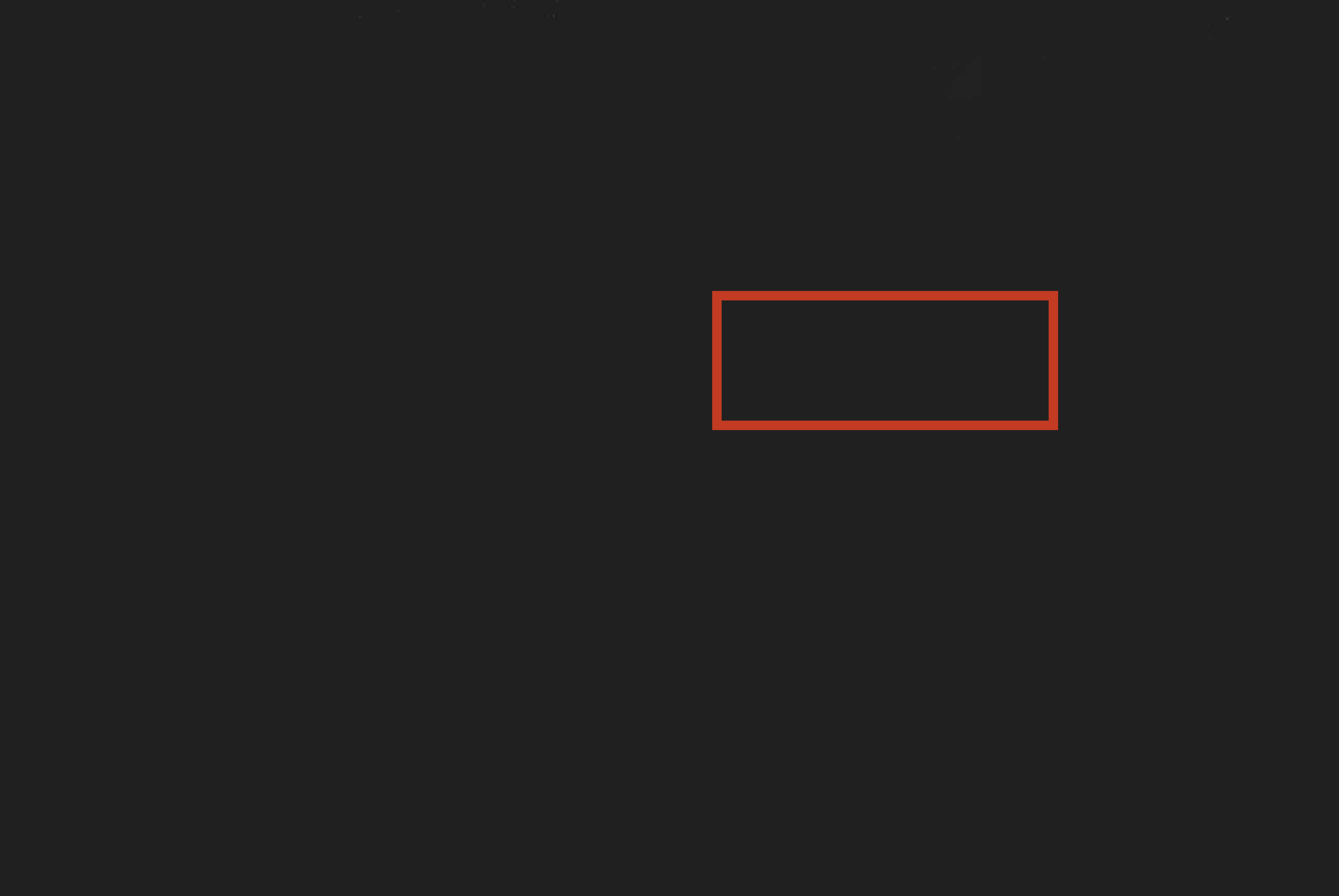}} &
         \frame{\includegraphics[width=0.23\textwidth]{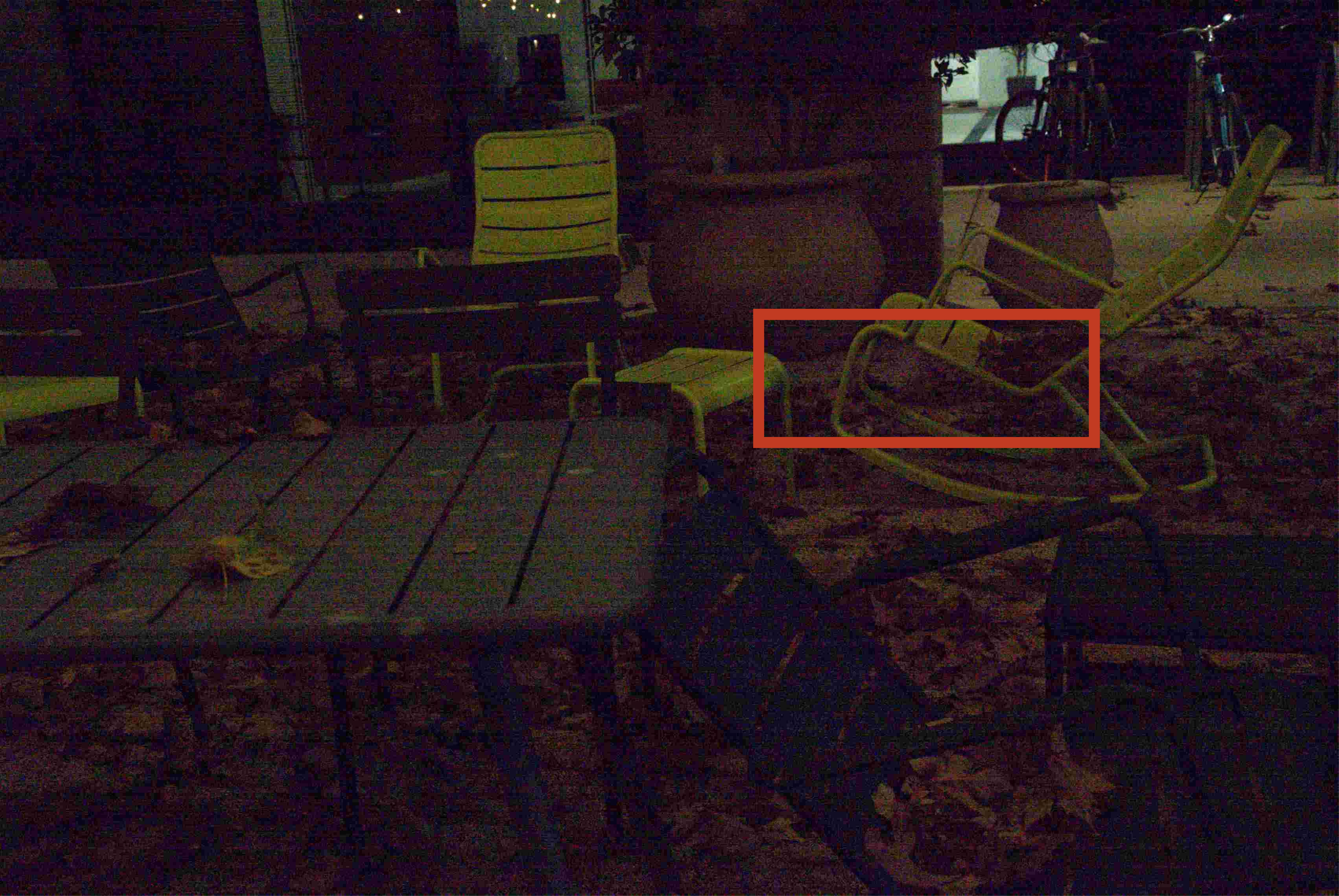}} & 
         \frame{\includegraphics[width=0.23\textwidth]{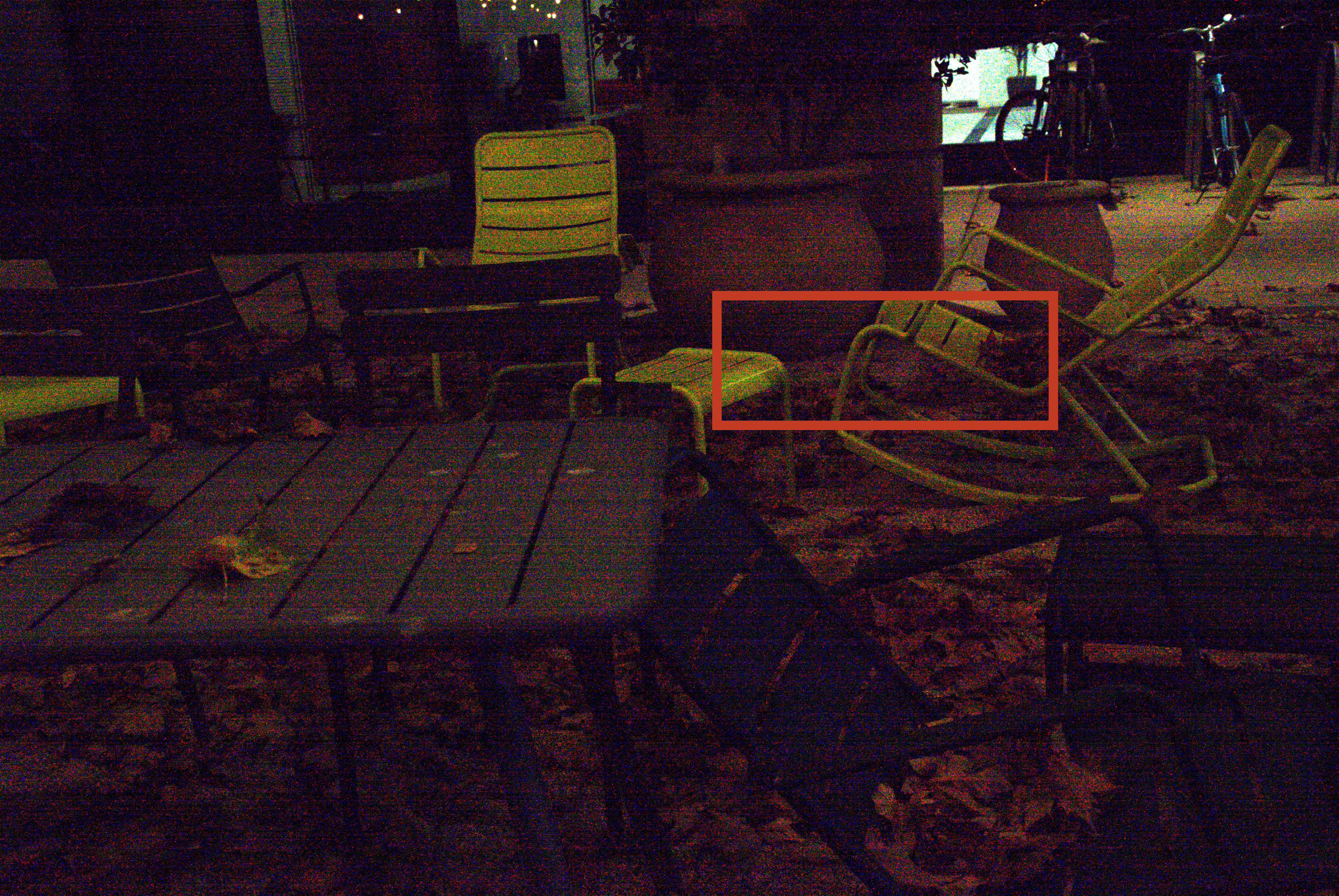}} & \frame{\includegraphics[width=0.23\textwidth]{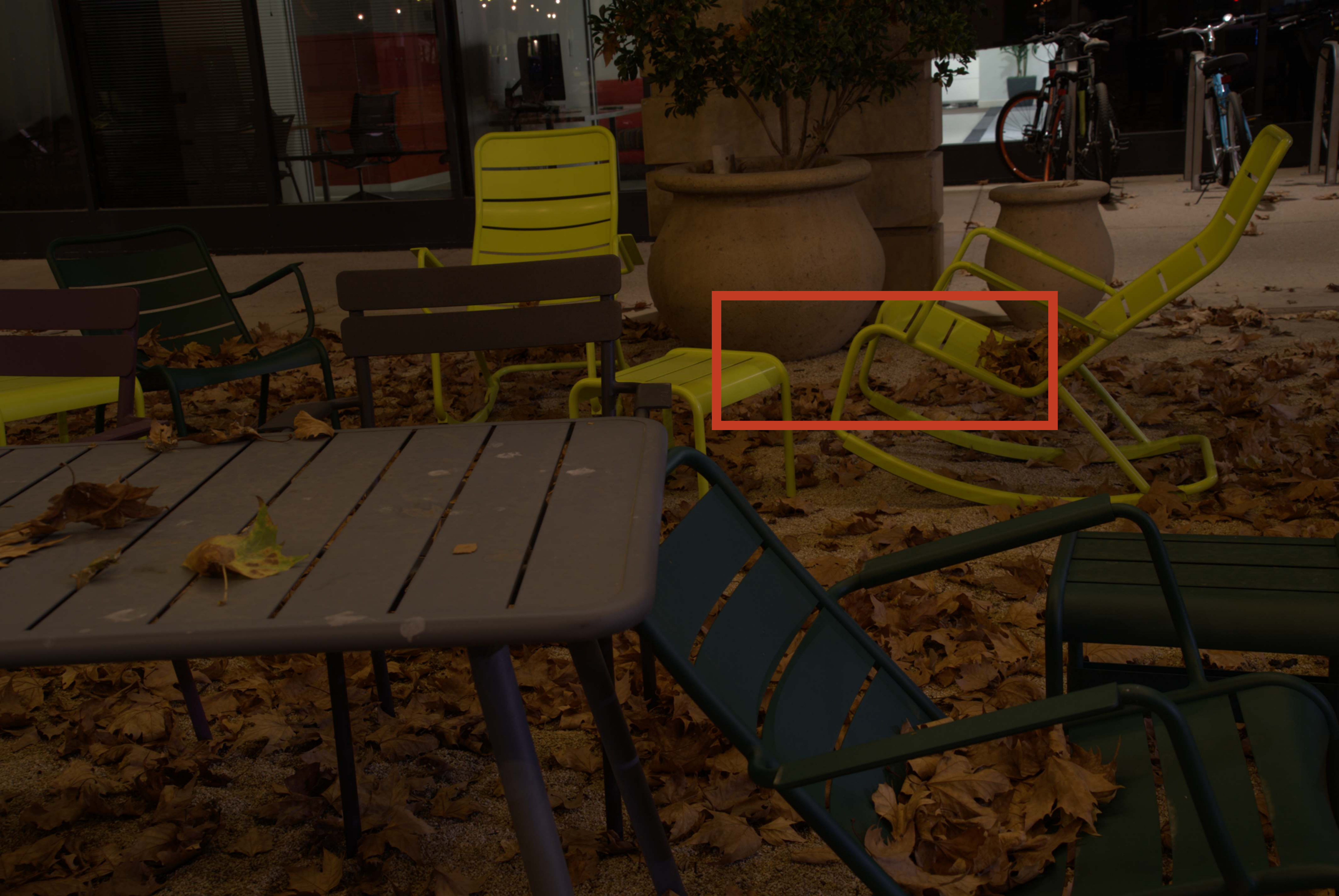}} \\
        & \includegraphics[width=0.23\textwidth, cfbox=red 1pt 0pt]{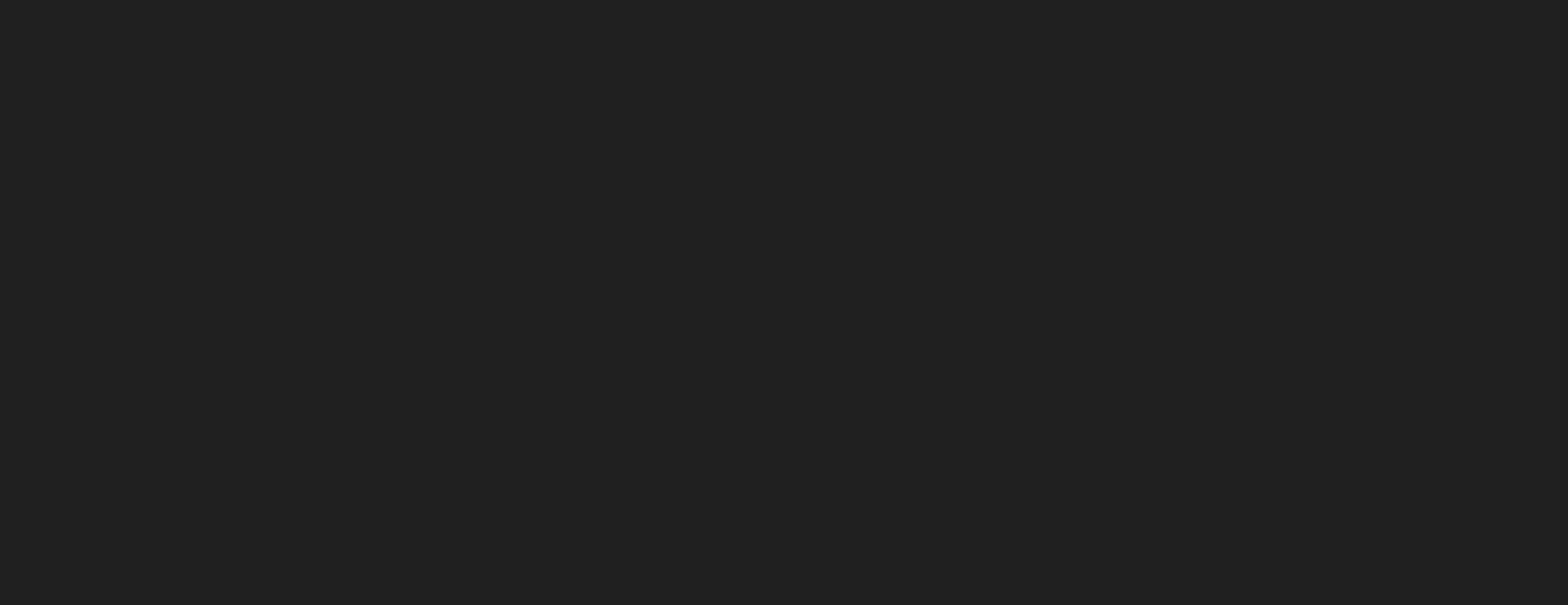} & \includegraphics[width=0.23\textwidth, cfbox=red 1pt 0pt]{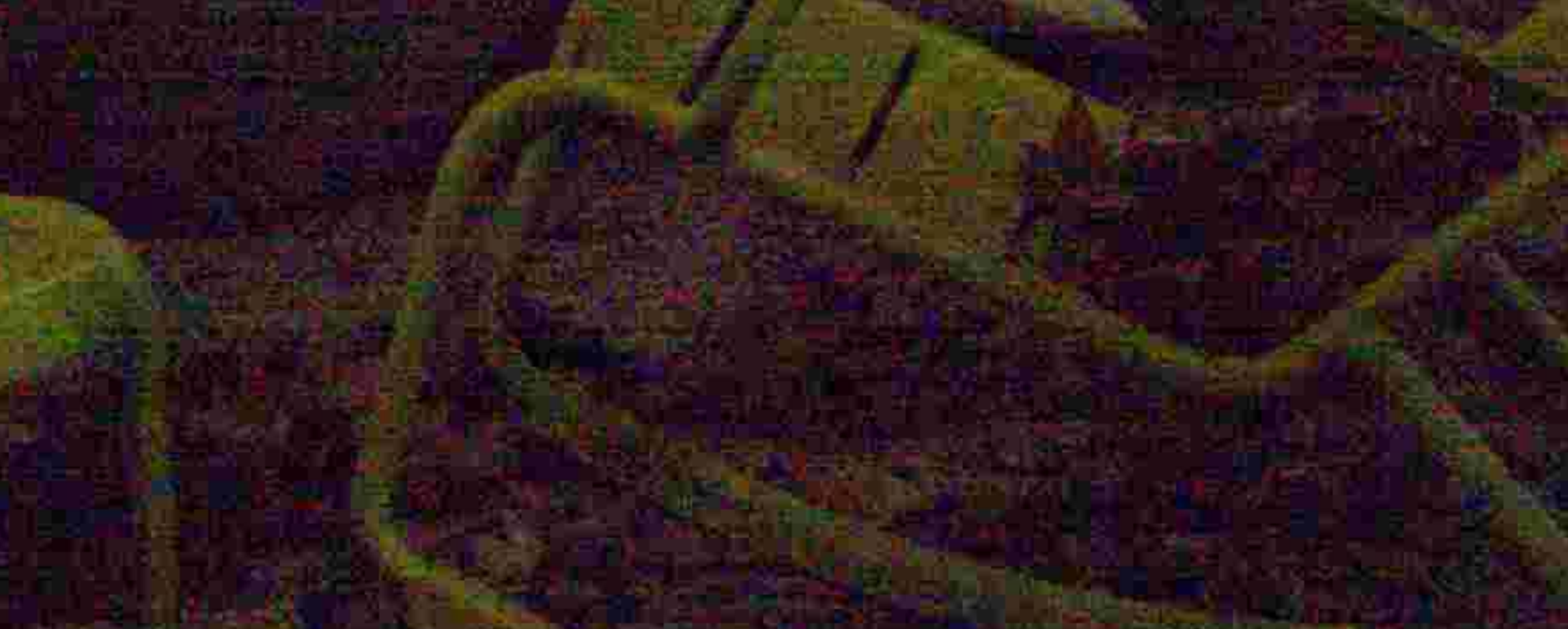} & \includegraphics[width=0.23\textwidth, cfbox=red 1pt 0pt]{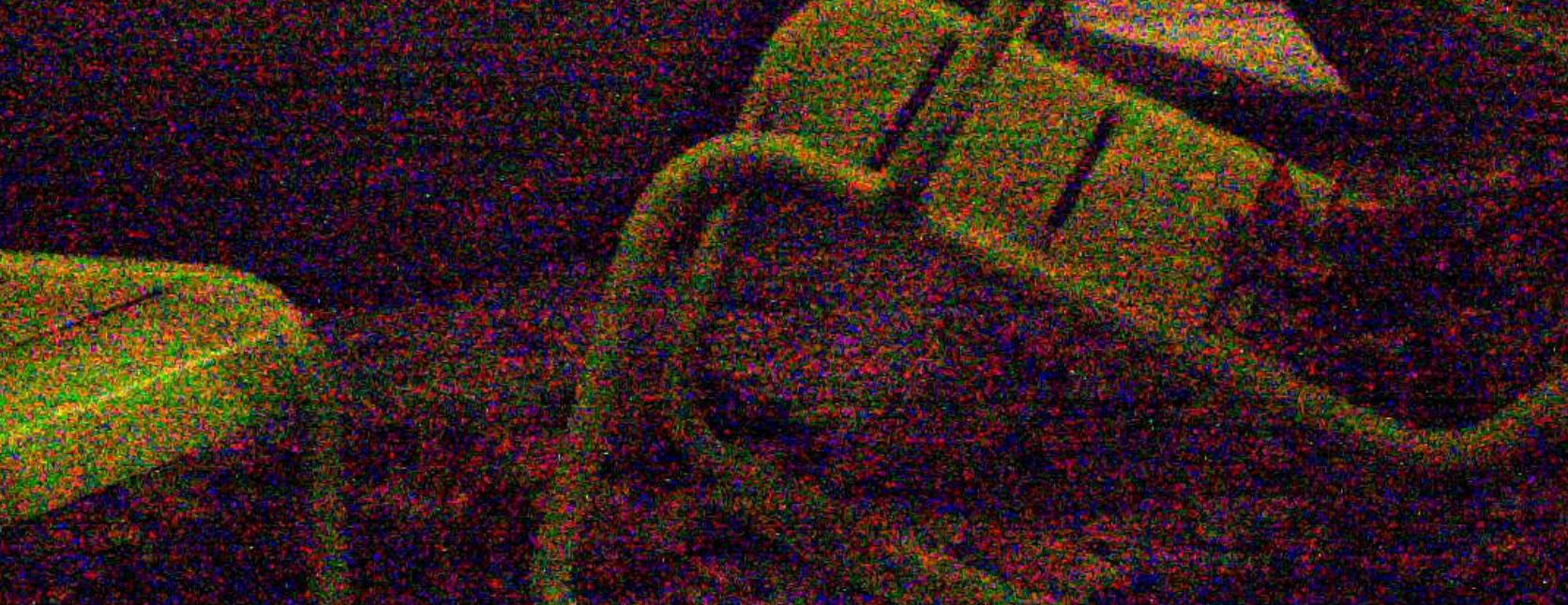} & {\includegraphics[width=0.23\textwidth, cfbox=red 1pt 0pt]{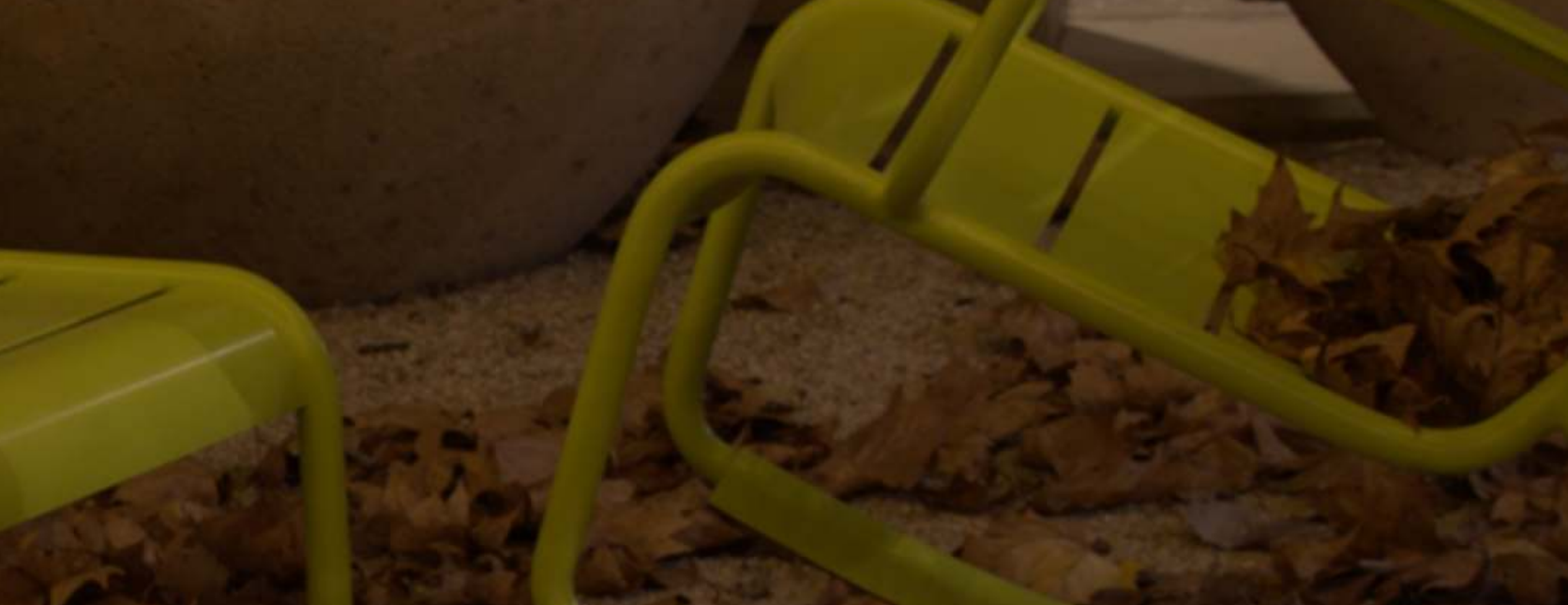}}\\ \\
         \multirow{2}{*}[12ex]{\rotatebox[origin=c]{90}{$\times300$, ISO12800 (indoor)}} &
         \frame{\includegraphics[width=0.23\textwidth]{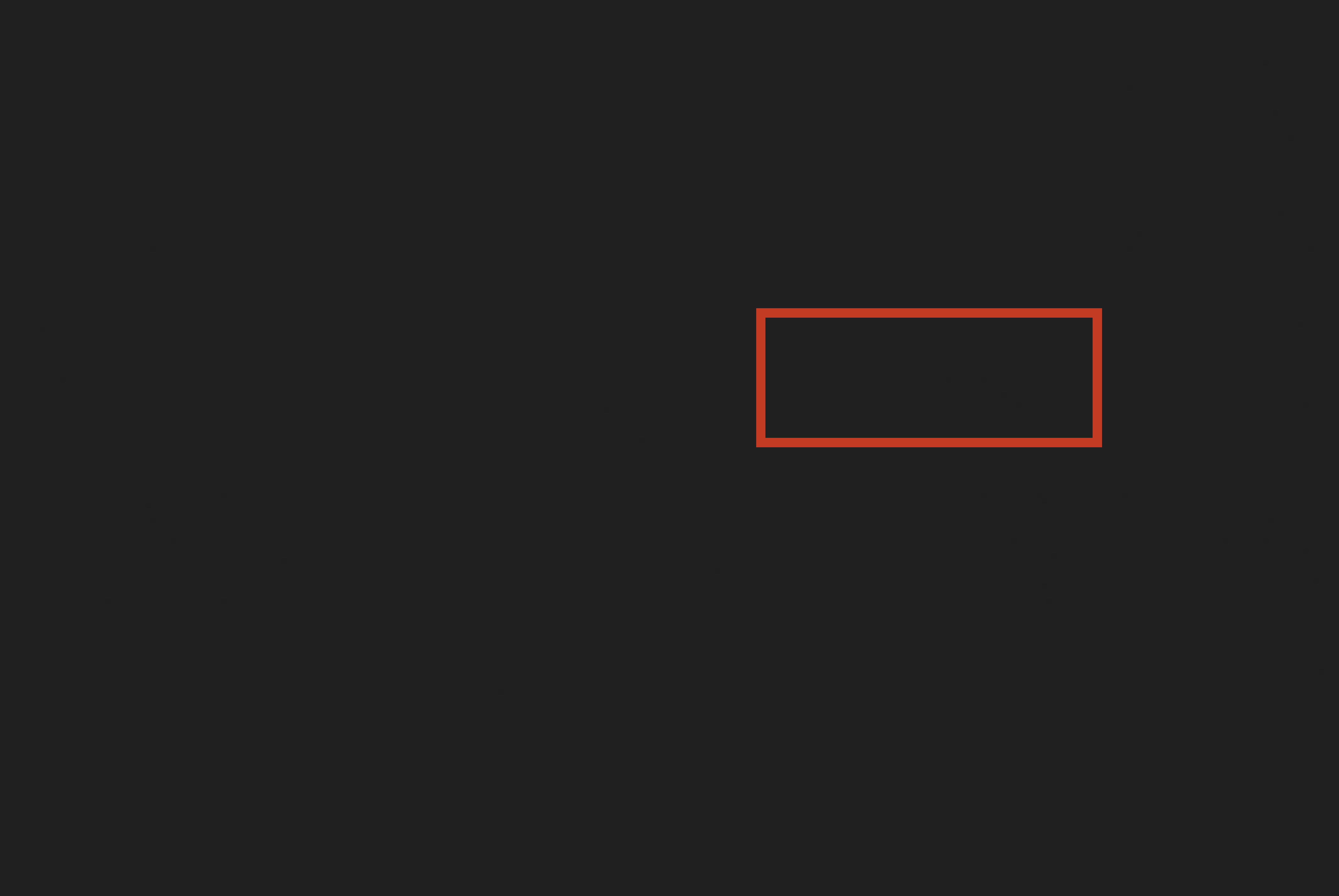}} & 
         \frame{\includegraphics[width=0.23\textwidth]{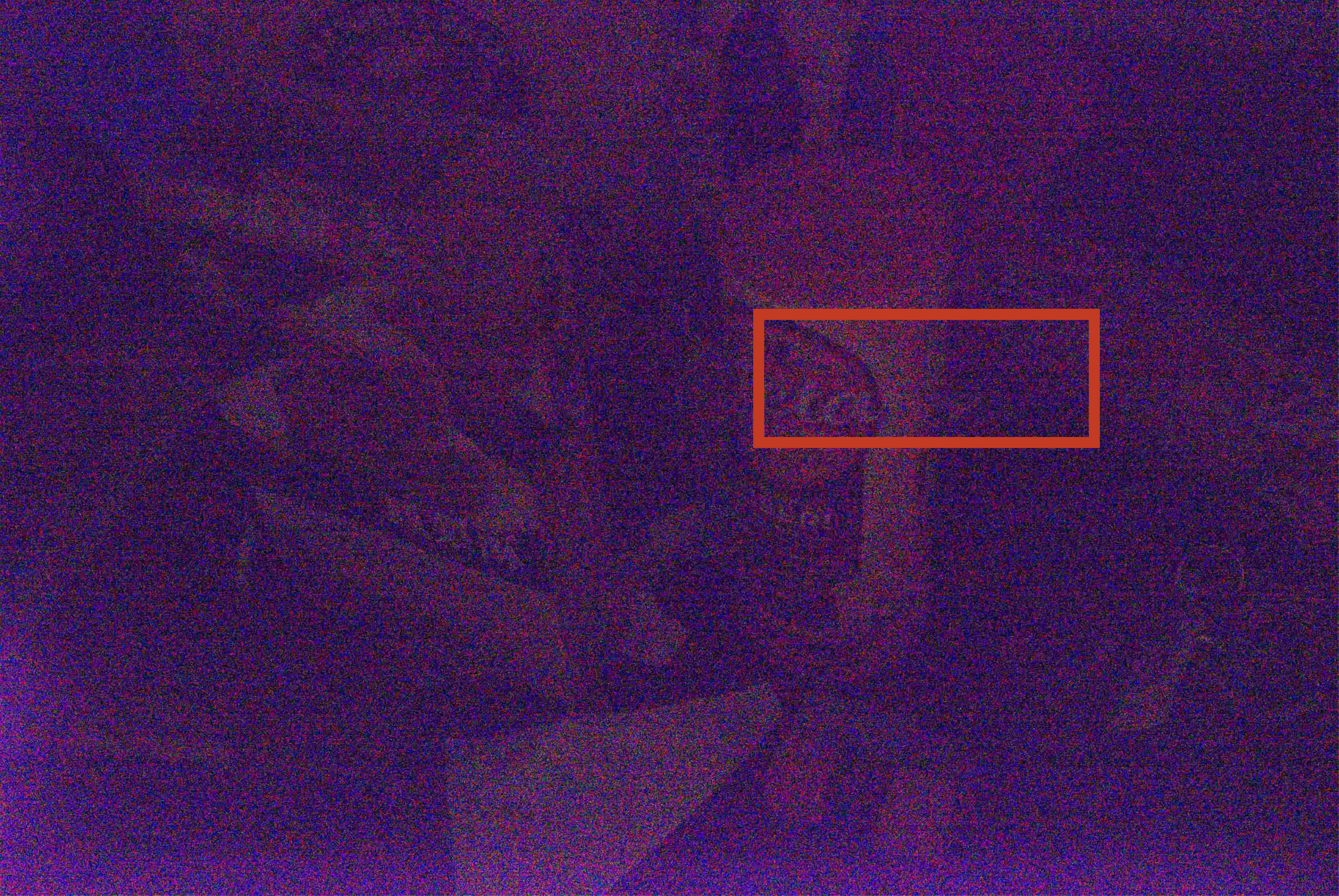}} & \frame{\includegraphics[width=0.23\textwidth]{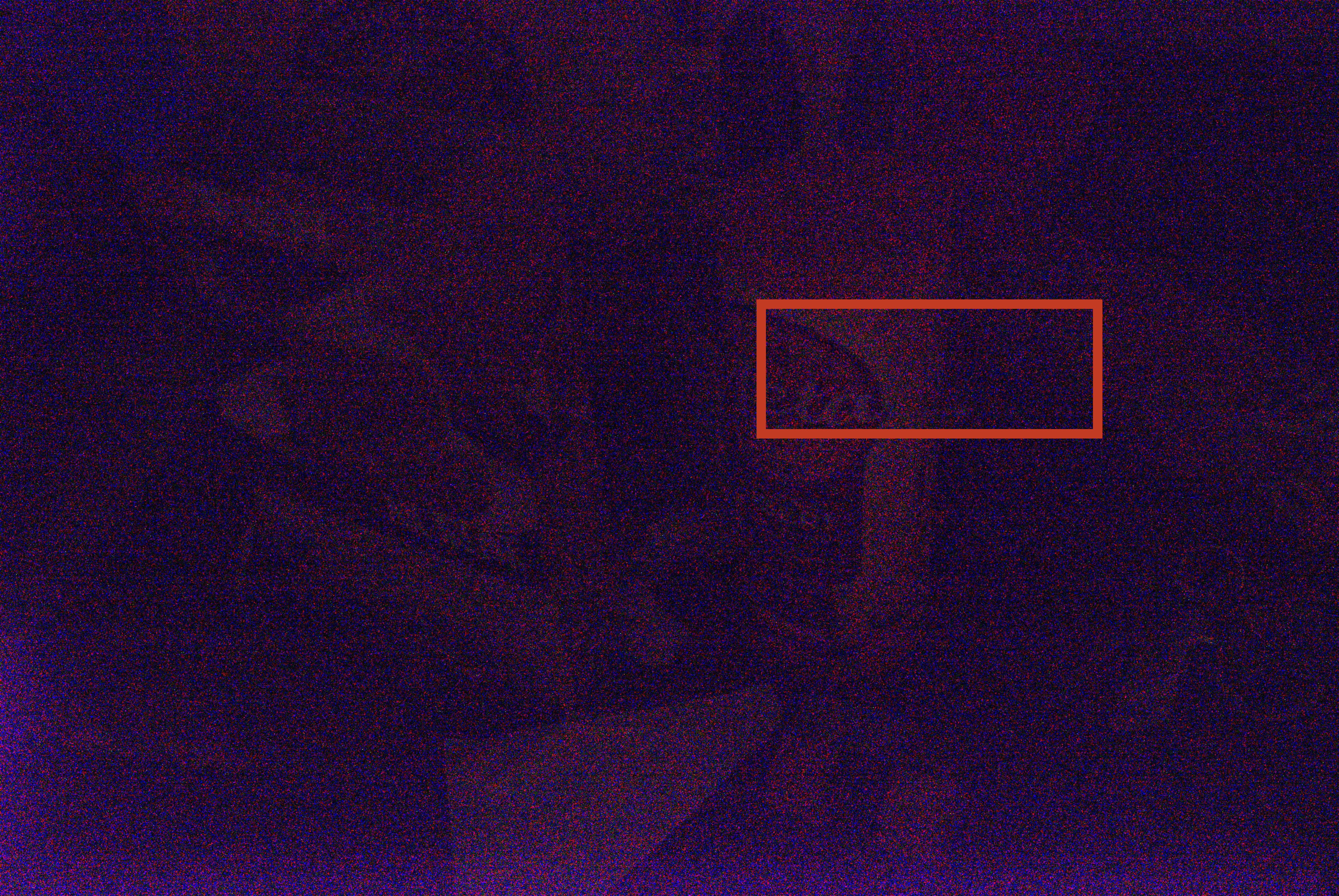}} & \frame{\includegraphics[width=0.23\textwidth]{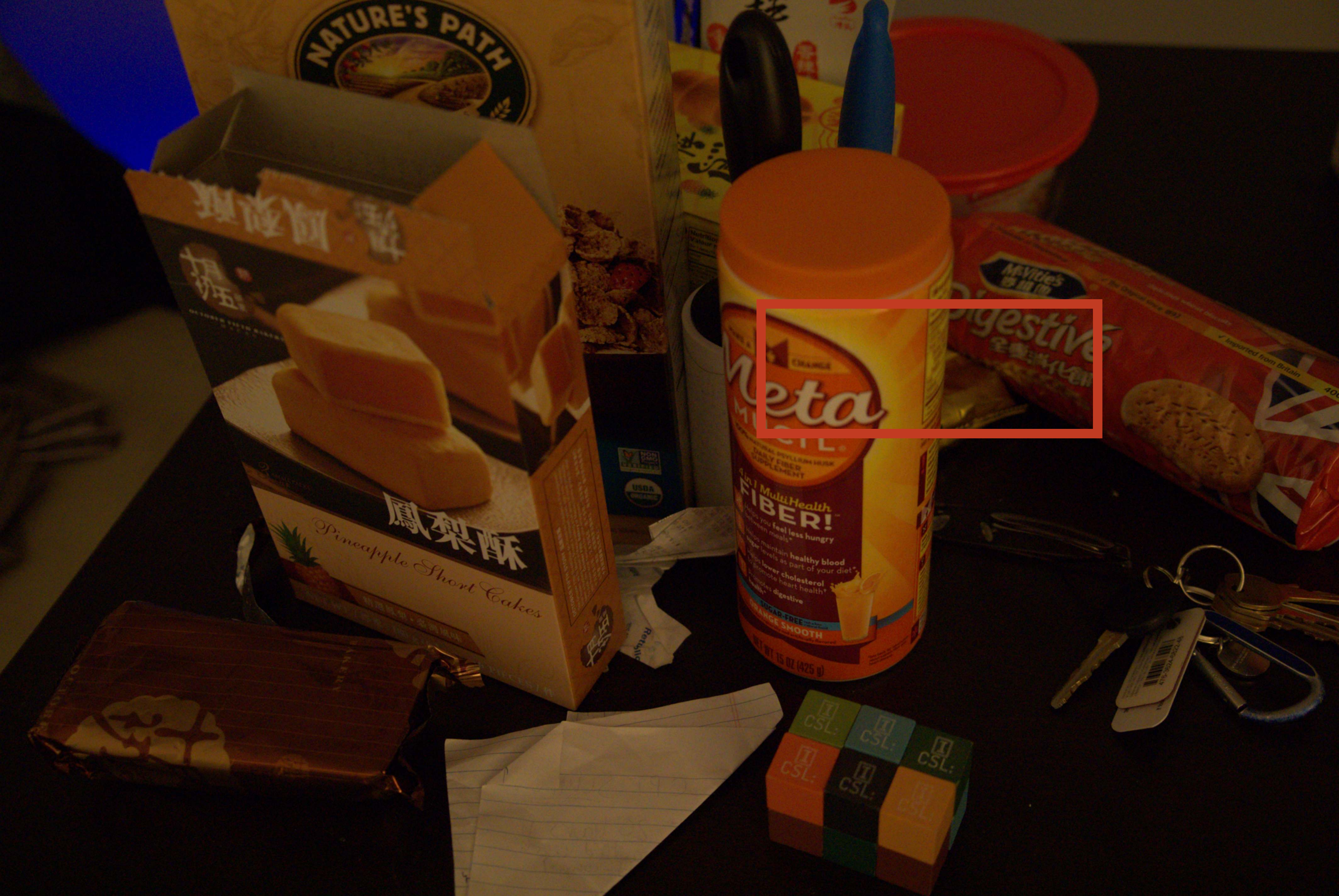}} \\
        & \includegraphics[width=0.23\textwidth, cfbox=red 1pt 0pt]{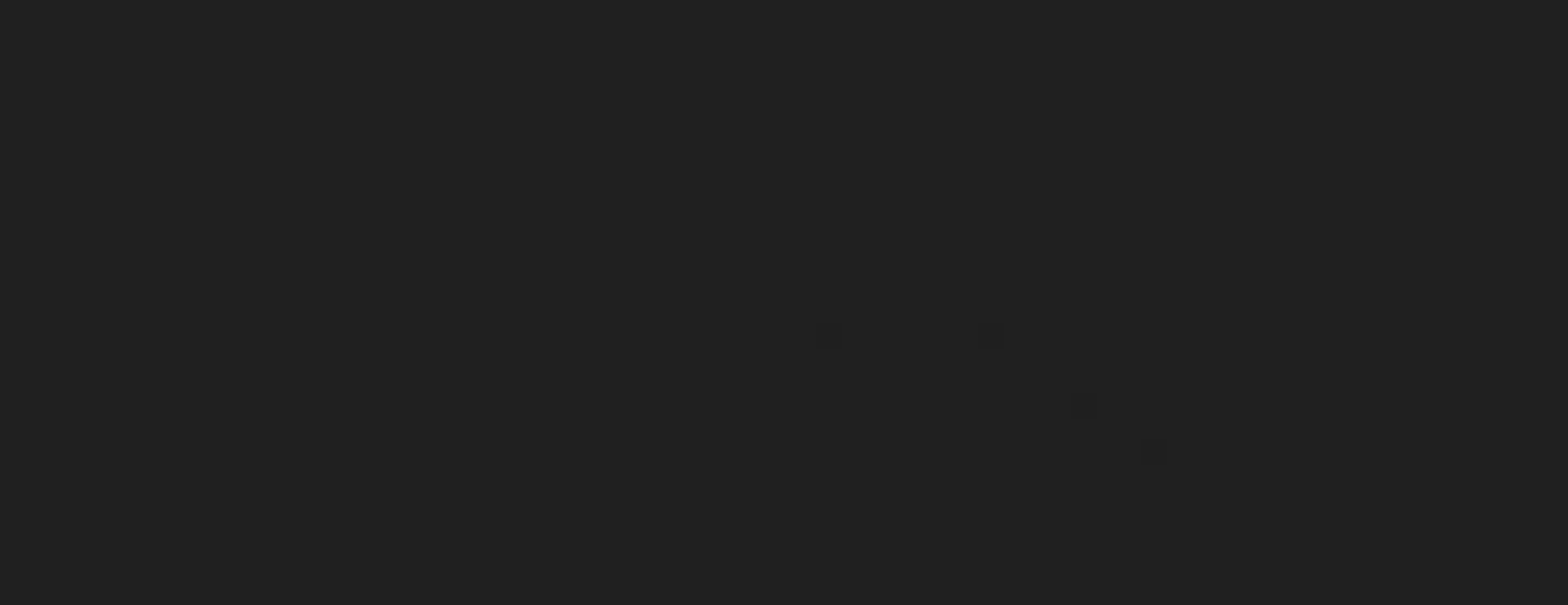} & \includegraphics[width=0.23\textwidth, cfbox=red 1pt 0pt]{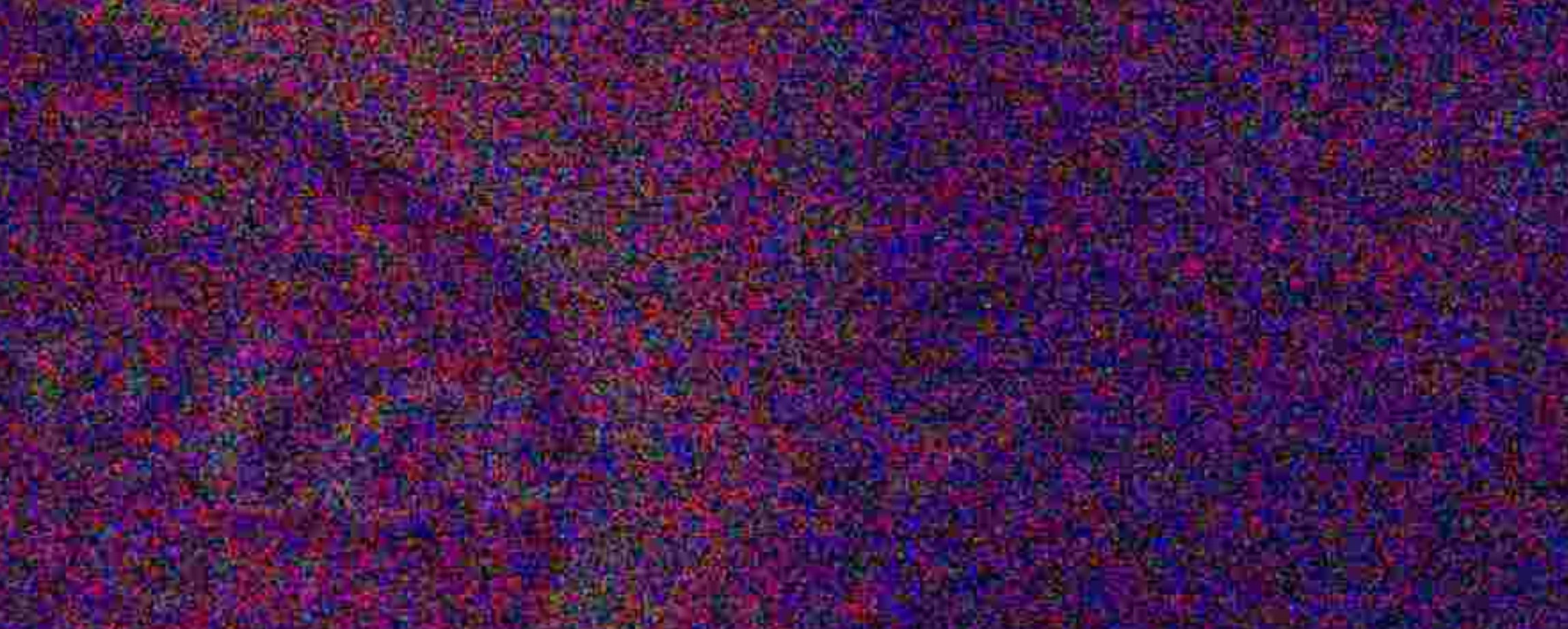} & \includegraphics[width=0.23\textwidth, cfbox=red 1pt 0pt]{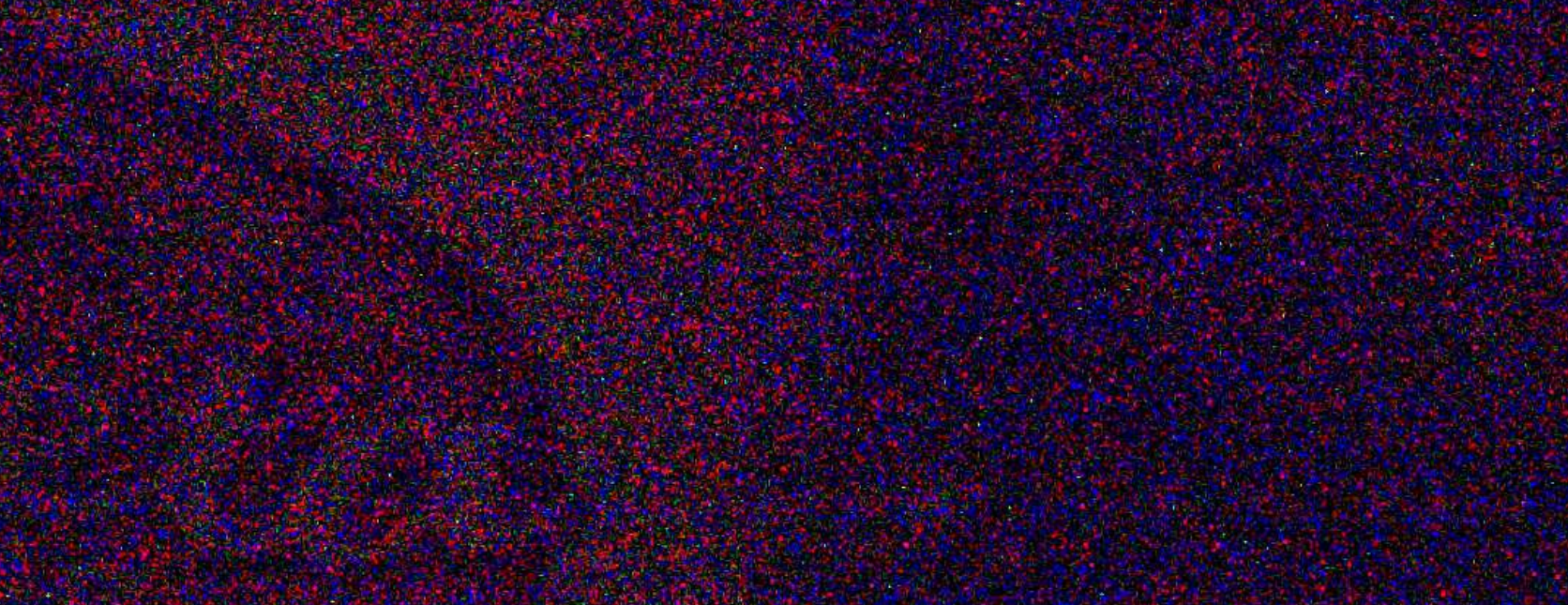} & {\includegraphics[width=0.23\textwidth, cfbox=red 1pt 0pt]{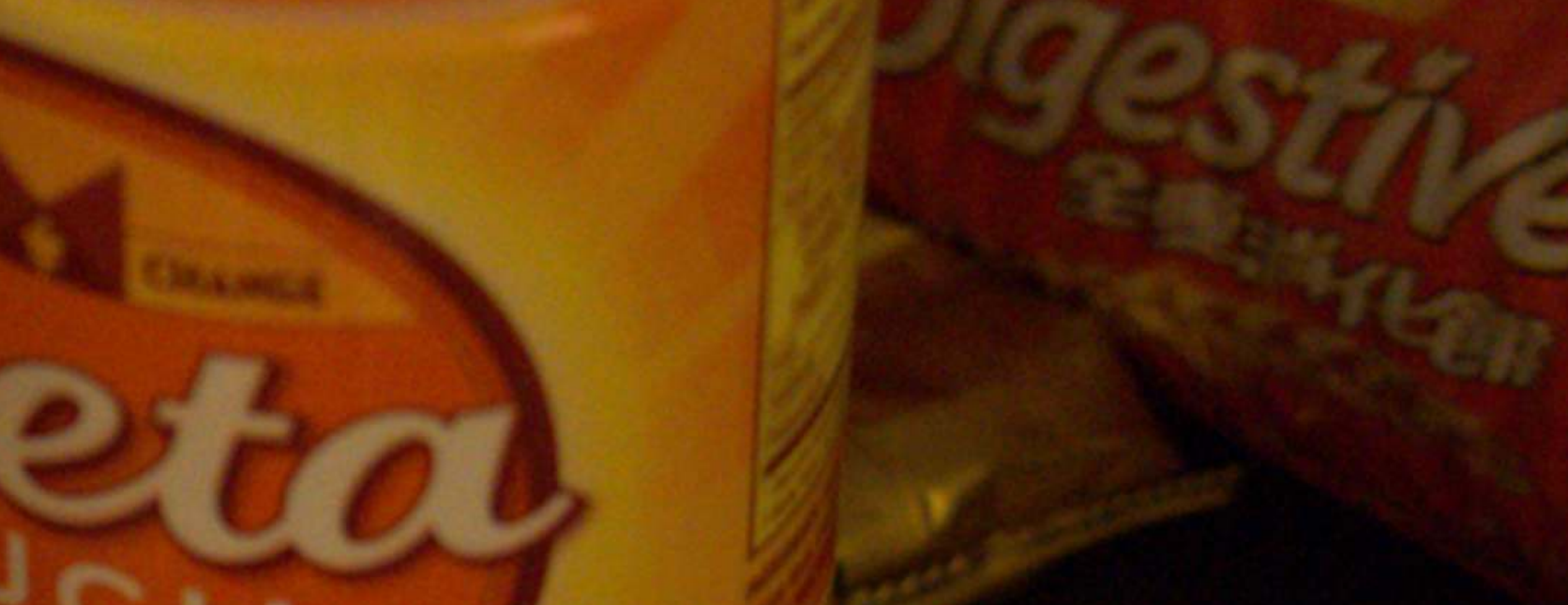}}
    \end{tabular}
    \captionof{figure}{Given a RAW image, \acronym \ can perform the entire image-processing pipeline for low-light enhancement. We show the camera output visualized in the sRGB space, the scaled-up RAW images, the output of a traditional pipeline, and compare them with our approach trained on the RAW data. These images are taken with Sony $\alpha$7SII and for these input images, we have that the illuminance at the camera is $<$ 0.1 lux, and we show the ratio of exposure times between reference ground-truth and input images ($\times250, \times300$). (\textbf{Best viewed in color and with zoom.})}
    \label{fig:teaser}
\end{center}%
}]
\maketitle

\begin{abstract}
Imaging under extremely low-light conditions presents a significant challenge and is an ill-posed problem due to the low signal-to-noise ratio (SNR) caused by minimal photon capture. Previously, diffusion models have been used for multiple kinds of generative tasks and image-to-image tasks, however, these models work as a post-processing step. These diffusion models are trained on processed images and learn on processed images. However, such approaches are often not well-suited for extremely low-light tasks. Unlike the task of low-light image enhancement or image-to-image enhancement, we tackle the task of learning the entire image-processing pipeline, from the RAW image to a processed image. For this task, a traditional image processing pipeline often consists of multiple specialized parts that are overly reliant on the downstream tasks. Unlike these, we develop a new generative ISP that relies on fine-tuning latent diffusion models on RAW images and generating processed long-exposure images which allows for the apt use of the priors from large text-to-image generation models. We evaluate our approach on popular end-to-end low-light datasets for which we see promising results and set a new SoTA on the See-in-Dark (SID) dataset. Furthermore, with this work, we hope to pave the way for more generative and diffusion-based image processing and other problems on RAW data.
\end{abstract}


\section{Introduction}
\label{sec:intro}

All imaging systems need to capture light to form an image. When very little light is present in the scene, imaging systems can only capture a few photons from the scene; when few photons are collected from the scene, the captured image is corrupted by noise sources. The main challenge in these low-light scenes is that the signal measured by the sensor is low relative to the noise inherent in the measurement process.

While low-light imaging is a longstanding challenge, traditional methods still struggle to restore images corrupted by noise. For instance, amplifying the measured analog or digital signals can result in the noise being amplified, and significantly degrades the quality of the image. Applying post-processing techniques such as scaling or histogram stretching may improve the image quality to some extent, but they cannot resolve the issue of low signal-to-noise ratio (SNR) due to the low photon counts in the captured image. Physical imaging techniques to increase SNR, such as opening the aperture or using flash, also have their own drawbacks, including motion blur and camera shake that make downstream applications with such images very challenging, and in many cases not feasible. We can also use burst-imaging techniques~\cite{10.1145/2980179.2980254}, however, these often fail in very low-light scenarios since these methods are reliant on robust alignment techniques to account for motion in the scene which is immensely challenging in the extreme noise setting in which we are interested in.

Alternatively, denoising algorithms can be applied to noisy images to improve image quality. These range from conventional techniques such as spatial filters like~\cite{tomasi1998bilateral, jain1989fundamentals, yang1995optimal} to other modern approaches like BM3D~\cite{4271520}, Deep Learning approaches based on CNNs~\cite{9756908, zamir2020learning, LORE2017650, Guo_2020_CVPR, Chen_2018_CVPR, 8784972}, and generative models based on: GANs~\cite{Zhu_2017_ICCV, 9334429}, normalizing flows~\cite{Wang_Wan_Yang_Li_Chau_Kot_2022}, and very recently on denoising diffusion~\cite{panagiotou2023denoising, nguyen2023diffusion, Monakhova_2022_CVPR}. Although these methods have been very successful in enhancing low-light images, these approaches do not target very low-light scenarios that we tackle or some of these approaches only tackle image enhancement as opposed to the entire image processing pipeline like this work. Though there have been approaches that tackle very low-light scenarios ($<0.1$ lux) like~\cite{7839189}, however, they use a simple CNN to solve this problem and we show how we propose a new generative image processing pipeline.
Applying our approach to captured RAW data outperforms other techniques that operate on pre-processed Low Dynamic Range (LDR) camera inputs,
and is able to correctly generate minute details of the image as well demonstrated by Figure~\ref{fig:teaser}.

Our approach specifically solves this problem by proposing a new image-processing pipeline based on a diffusion model. We first use a standard pre-trained latent diffusion model and fine-tune this model on patches of up-sampled and corrected Bayer RAW images. The fine-tuned model then learns the entire image processing pipeline for low-light RAW data including but not limited to demosaicing, color correction, enhancement, and transformations. The fine-tuning process itself is supervised by long-exposure ground-truth images and the model is fine-tuned to produce the same resolution enhanced images. This greatly improves the overall image-processing pipeline for low-light RAW data and reduces the use of handcrafted processes in the image-processing pipeline, over other traditional and learned image-processing pipelines. Furthermore, this approach also easily allows combining our learned model with other forms of popular text conditioning like Dreambooth~\cite{ruiz2023dreambooth} and Instruct fine-tuning~\cite{brooks2023instructpix2pix} among others and can potentially be extended to support other high-level image editing operations like stylization, inpainting, accessorization, and property modification. This allows our approach also enables a new type of end-to-end generative camera Image Signal Processor (ISP) wherein regions of an image can be inpainted or stylized using models that are applied end-to-end, starting from the RAW data.

Our approach not only demonstrates improvements throughout the image processing pipeline for low-light RAW data but also is the first approach using a diffusion-based model for working with RAW images and makes apt use of the vast information a RAW image has for the downstream task, here enhancing and denoising very low-light images. With this paper, we hope to pave the way for more diffusion-based approaches for a generative end-to-end camera processing pipeline leading to significant performance improvements.

\paragraph{Contributions.} A traditional image processing pipeline often consists of multiple specialized parts that are overly reliant on the downstream tasks. The key novelty of our approach stems from modifying the pipeline with a generalizable data-driven approach.
\begin{itemize}
    \item We propose a first-of-its-kind generative camera ISP for extreme low-light image enhancement based on a diffusion model that improves performance on low-light image enhancement in most cases and is the first diffusion-based model that is modeled over RAW image.
    \item We develop an implementation of this technique using a latent diffusion model fine-tuned on low-light RAW images allowing for great extensibility and making use of the vast amount of priors that diffusion models have.
    \item We set a new state-of-the-art for image denoising on the popular See-in-the-Dark (SID) dataset~\cite{Chen_2018_CVPR} demonstrating that a fine-tuned diffusion model can enhance extremely low-light RAW images and perform the whole image processing pipeline on such data.
\end{itemize}

\section{Related Work}

Improving low-light images is a problem that has been very well explored throughout the years, and many deep-learning-based methods have recently performed very well in improving low-light images which has led to it being practical now. Images taken in low light have a low signal-to-noise ratio in the captured sensor data and thus are very noisy, to improve such low-light images, these images need to be denoised. We provide a brief overview of image-denoising algorithms. Our approach also makes use of latent diffusion models and we also provide a brief overview of diffusion models. A popular class of algorithms for improving low-light images are ones that mainly handle processed images or image enhancement algorithms as well as algorithms that handle the end-to-end image processing pipelines for this problem. We also provide a brief overview of both of these classes of algorithms.

\subsection{Image Denoising}

Image denoising is a very well-explored topic in vision and there have been many approaches proposed for denoising images. Many traditional image-denoising approaches draw inspiration from various mathematical and machine learning techniques, utilizing image priors, statistical insights, and the power of deep learning architectures. Within classical image denoising techniques, methods such as Total Variation (TV) denoising~\cite{rudin1992nonlinear}, wavelet-domain processing~\cite{portilla2003image}, sparse coding~\cite{elad2006image}, nuclear norm minimization~\cite{gu2014weighted}, and 3D Transform-Domain Filtering (BM3D)~\cite{dabov2007image} perform well across a range of noise levels and exploit specific priors such as sparsity intensity changes, and transformation domain.

The advent of deep learning approaches has significantly reshaped this landscape. There have been multiple convolution-based denoising approaches~\cite{YIN2020107717, liu2021adnet}. These approaches learn convolutional networks to approximate the noise maps and are often based on strong priors. Furthermore, there have also been other specialized approaches for denoising like Stacked Sparse Denoising Auto-Encoders (SSDA)~\cite{wang2004image}, Trainable Nonlinear Reaction-Diffusion (TNRD)~\cite{chen2016trainable}, and Deep Autoencoders~\cite{lore2017llnet} demonstrating performance increases over traditional methods in capturing intricate noise patterns and restoring images.~\cite{Monakhova_2022_CVPR} proposed using physics-inspired GANs with stochastic corruptions during training, to enhance the appearance of night sky images.

\subsection{Diffusion Models}

Diffusion Models~\cite{NEURIPS2020_4c5bcfec} are a type of probabilistic generative model that has gained popularity in recent years due to its ability to generate high-resolution images with diverse features. The underlying principle behind diffusion models involves a forward process that gradually adds noise to clean samples drawn from a prior distribution and a reverse process that reverses the corruption process to recover plausible samples from the noise~\cite{luo2022understanding}. Diffusion models have demonstrated their effectiveness in various image-based tasks, such as unconditional image generation, inpainting, colorization, image segmentation, and medical imaging. Compared to other generative models such as GANs and VAEs, diffusion models offer a stable training process and the ability to learn strong priors \cite{NEURIPS2020_4c5bcfec}. A Diffusion Model generates samples by gradually removing noise from a signal and learning a network to do so, while their training objective is a reweighted variational lower bound, have shown particularly great results for denoising images as well. In this work, we work on fine-tuning the broad class of latent diffusion models~\cite{rombach2022high}.

Another recent approach to low-light image enhancement using deep learning is based on probabilistic models, such as conditional denoising diffusion probabilistic models (DDPMs). Diffusion in the Dark (DiD)~\cite{nguyen2023diffusion} is a recent diffusion model for low-light image reconstruction that provides qualitatively competitive reconstructions, especially for low-light text recognition. In contrast, our approach relies on fine-tuning a latent Diffusion model directly on RAW sensor data and aims at learning the entire image processing pipeline.

\begin{figure}[t]
\centering
\includegraphics[width=\columnwidth]{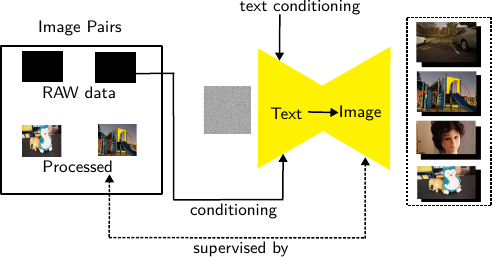}
\caption{We show how our model is fine-tuned on text-to-image diffusion models which requires paired RAW sensor data and processed long-exposure image.}
\label{fig:in-out}
\end{figure}
\begin{figure*}[!ht]
\centering
\includegraphics[width=\textwidth]{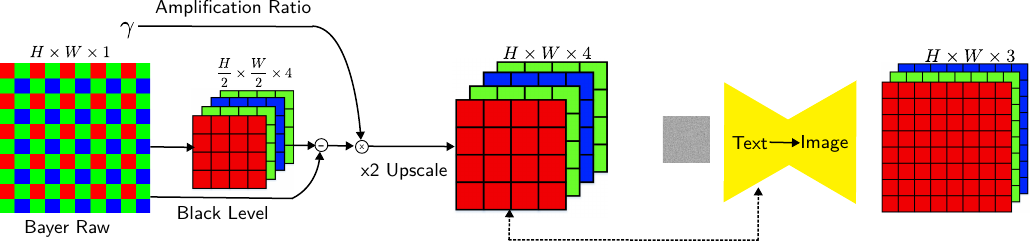}
\caption{Our fine-tuning approach consists of first naively packing the RAW images, subtracting the black level, scaling the image by an externally set amplification factor, upscaling the packed RAW images, and then fine-tuning a text-to-image latent diffusion model conditioned on these packed RAW images.}
\label{fig:pipeline}
\end{figure*}

\subsection{Low-light Photography.} Classical low-light reconstruction methods have been widely used to enhance the quality of low-light images, and they remain an active area of research. These methods can be broadly classified into histogram equalization-based methods~\cite{CHENG2004158, 4429280, 4266947, kaur2011survey}, Retinex-based methods~\cite{597272, 10.1117/1.1635366, 4130414, 4483799, 7229296, 7782813}, and burst averaging techniques~\cite{Delbracio_2015_CVPR, Aittala_2018_ECCV, 10.1007/978-3-319-54190-7_3}.

Histogram equalization, as described in\cite{CHENG2004158}, aims to enhance image contrast by redistributing pixel intensities. However, it often suffers from over-enhancing noise in darker regions. In contrast, Retinex-based methods separate images into reflectance and illumination components, inspired by Land's theory of color vision~\cite{land1977retinex}. These methods include single-scale Retinex (SSR)\cite{7080567, kaur2011survey}, multi-scale Retinex (MSR)~\cite{rahman1996multi}, and adaptive MS-Retinex (AMS-R)~\cite{lee2013adaptive, lin2014multi}, each with its unique advantages.

To address the limitations of classical techniques, recent research has turned to deep learning approaches. Convolutional neural networks (CNNs) have been employed in various papers such as \cite{Ignatov_2019_CVPR_Workshops} and \cite{zamir2020learning}, autoencoders in \cite{bigdeli2017image} and \cite{Shimobaba:17}, and generative adversarial networks (GANs) in \cite{Ignatov_2018_CVPR_Workshops}, and \cite{Monakhova_2022_CVPR}. Additionally, diffusion-based methods have been explored, as seen in the work by~\cite{nguyen2023diffusion}. These modern techniques enhance classical methods by decoupling illumination and reflectance, utilizing attention mechanisms, or applying normalizing flows to achieve improved performance and noise robustness. Some deep learning methods even operate without the need for paired low-light/well-lit image data, enhancing their practicality for real-world applications~\cite{Liang_2021_ICCV, Zamir_2022_CVPR, 10.1007/978-3-031-25063-7_21, NEURIPS2022_3ca6d336}.

However, image enhancement algorithms are often not suitable to be used as an entire image processing pipeline for enhancing low-light images. This is often because most image enhancement algorithms are only trained with processed image pairs, which causes such methods to be unable to perform the multitude of steps required for the image processing pipeline as well as causes these algorithms to not be able to handle very low-light images.

\subsection{Learning-based image processing pipeline}

Recently there have been multiple data-driven or learned image-processing pipelines developed to improve low-light images. These approaches learn image-processing pipelines using hand-crafted features~\cite{wei1993real} or deep-learning approaches~\cite{Morawski_2022_CVPR, ZAMIR202137, Chen_2018_CVPR}. These algorithms aim to propose a new image processing pipeline directly on RAW images and the corresponding produced images are enhanced.

There have been approaches that train models to learn the entire image-processing pipeline~\cite{Chen_2018_CVPR, ZAMIR202137, Wei_2020_CVPR, 10.1145/3503161.3548186, Wang_2023_ICCV, Zhang_2023_ICCV}. These works usually train a model that takes a RAW image or a lightly processed RAW image and handles the entire pipeline of creating an image while performing some enhancement like producing denoised images from low-light settings. Our work is closest to LSID~\cite{Chen_2018_CVPR} which tackles the same problem, of learning an image processing pipeline for extremely low-light images. LSID tackles this problem by learning a convolutional network and handcrafting some features like the brightness of the generated image or the black level, this is trained on image pairs of scenes with high exposure time and the same scene with varying levels of lower exposure time, which generates noisy images. The convolution network not only learns the entire image processing pipeline but also denoises this image well. However, our work develops a new generative camera-processing pipeline contrary to LSID. Our approach is based on fine-tuning existing diffusion models to learn this image-processing pipeline which improves upon the performance of LSID and other subsequent methods and is also the first work to present a diffusion model fine-tuned on RAW images. Our approach is also different from other Diffusion-based approaches~\cite{nguyen2023diffusion, panagiotou2023denoising} which only perform image enhancement on post-processed images, unlike our approach which is a new end-to-end generative camera processing pipeline.

\section{Method}

Given a casually captured RAW image in very low-light settings, without any description or any other information about what the scene contains, our objective is to generate a processed final image with the scene as if it was taken in higher-light settings with more exposure. We do not impose any restrictions on input image capture settings and the captured images can have varying numbers of photons captured per pixel. Our approach being based on fine-tuning, in the future can easily be modified to be guided by more diverse text and support other high-level image editing operations like stylization, inpainting, accessorization, and property modification. We next provide some background on conditional latent diffusion models (Sec. \ref{sec:conditionallatentmodels}), and then present our fine-tuning technique to develop a generative ISP (Sec. \ref{sec:pipeline}).

\subsection{Conditional Latent Diffusion Models}
\label{sec:conditionallatentmodels}

Latent Diffusion models \cite{rombach2022high} project the input into a lower-dimensional latent space and train the diffusion model on this latent space instead of applying the diffusion process on a high-dimensional input pixel space. Specifically, the forward and backward diffusion processes occur in a lower-dimensional latent space and an encoder-decoder architecture is trained on a large image dataset to translate images into latent codes. At inference time, a random noise latent code goes through the backward diffusion process, and the pre-trained decoder is used to generate the final image. We have a conditional latent diffusion model $\bepsilon_\theta$, which can be interpreted as a sequence of denoising autoencoder $\bepsilon_\theta(\bz_t, t, \ldots)$; $t \sim \mathcal{U}(\{1, \ldots, T\})$, that is trained using a squared error loss (or represent the denoising term in the ELBO) to denoise some variably noisy images as follows:
\begin{equation}
\Eb{\mathcal{E}(\bx),\bc,\bepsilon \sim \mathcal{N}(0,I),t}{\| \bepsilon - \bepsilon_{\theta}(\bz_t, t, \tau_\theta(\bc)) \|^2_2}
\end{equation}
where $\bx_t$ is some noisy image, $\bc$ is a conditioning vector, $\tau_\theta$ is an encoder that projects $\bc$ to an intermediate representation, $\mathcal{E}$ encodes $\bx_t$ into a latent representation $\bz_t$, and we jointly optimize $\bepsilon_\theta$ and $\tau_\theta$.

\subsection{\acronym: Generative ISP}
\label{sec:pipeline}

In typical image processing pipelines, RAW data obtained from an imaging sensor undergoes several sequential module applications, including white balance, demosaicing, denoising, sharpening, color space conversion, gamma correction, etc., where these modules are frequently calibrated specifically for individual camera models. \cite{schwartz2018deepisp} suggested using numerous localized, linear, and learned filters collectively termed ``L3'' filters to mimic the intricate nonlinear pathways witnessed in current-day commercial imaging solutions.
Previous deep-learned ISP methods are often trained using some kind of convolutional architecture~\cite{Chen_2018_CVPR, schwartz2018deepisp}. However, processing low-light images is an ill-posed problem since many possible reconstructed images could be consistent with the measurements. Using pre-trained generative models allows us to exploit learned priors and generate processed images that obey the learned natural image statistics.
Our approach needs to learn these multiple stages of a typical image processing pipeline for it to be an end-to-end generative ISP. This pipeline is summarized in Figure \ref{fig:pipeline}.

\paragraph{Demosaicing.} Demosaicing is a crucial step in image processing pipelines, particularly in digital imaging systems that use color filter arrays to capture images. In such systems, each pixel on the image sensor is assigned to detect only one primary color resulting in a spatially undersampled color image. Demosaicing reconstructs a full-color image from this mosaic by estimating the missing color values at each pixel location.

The sensor first gives us a Bayer RAW image and then our approach involves compressing the given information naively into four distinct channels. Consequently, there is a reduction in the original spatial resolution by half in both dimensions, $\left(\frac{H}{2}, \frac{W}{2}\right)$-sized images. 
Additionally, we then subtract the black level and scale the brightness of the input images by some amplification factor, $\alpha$ that is manually set. We then employ simple bicubic upsampling to upsample the input image back to $\left(H, W\right)$-sized images

\paragraph{Fine-tuning Approach.}

Incorporating a conditioning vector in classifier-free diffusion guidance~\cite{ho2021classifierfree} relies on training a conditional and an unconditional model for some conditioning $\bc$. This score estimate is modified towards the conditional model and away from the unconditional model. So we can calculate the modified score estimate as follows:
\begin{equation}
\begin{split}
    \Tilde{e_\theta}(\bz_t, \bc) &= \underbrace{e_\theta(\bz_t, \phi)}_{\substack{\text{unconditional model}}} \\&+ s \underbrace{(\underbrace{e_\theta(\bz_t, \bc)}_{\substack{\text{conditional}\\ \text{model}}}-\underbrace{e_\theta(\bz_t, \phi)}_{\substack{\text{unconditional}\\ \text{model}}})}_{\substack{\text{move towards conditional model}\\ \text{and away from unconditional model}}}
\end{split}
\end{equation}
where $s$ represents the level of guidance, $\bc$ represents the conditioning vector, $\phi$ represents a conditioning vector representing some fixed null value, and $\bz_t$ represents the latent representation.

Following this, in our case, we have two conditioning vectors: the input dark images $\bc_I$ and the text conditioning vector for high-level editing tasks $\bc_T$. We use classifier-free diffusion guidance for both of these conditioning vectors and compose score estimates from these two conditioning vectors~\cite{10.1007/978-3-031-19790-1_26}. During fine-tuning, we randomly set 
only $\bc_T=\phi_T$ for $5\%$ of examples, and both $\bc_I=\phi_I$ and $\bc_T=\phi_T$ for $5\%$ of examples which allows the model to be capable of conditional or unconditional denoising with respect to both or any of conditional inputs: on RAW images or text conditioning. As in InstructPix2Pix~\cite{10.1007/978-3-031-19790-1_26}, we also have two separate guidance scales $s_I$ representing the image guidance scale, and $s_T$ representing the text guidance scale, which can be adjusted to trade off how strongly the generated samples correspond with the input RAW image and how strongly they correspond with the text conditioning. Thus, our score estimate is as follows:

\begin{equation}
    \begin{split}
        \Tilde{e_\theta}(\bz_t,\bc_I, \bc_t) &= \underbrace{e_\theta(\bz_t, \phi_I, \phi_T)}_{\text{unconditional model}}\\
        &+ s_I \underbrace{(\underbrace{e_\theta(\bz_t, \bc_I, \phi_T)}_{\substack{\text{image conditioned}}} - \underbrace{e_\theta(\bz_t, \phi_I, \phi_T)}_{\text{unconditional model}})}_{\substack{\text{move towards $\bc_I$ conditioned model} \\ \text{and away from unconditional model}}}\\
        &+ s_T \underbrace{( \underbrace{e_\theta(\bz_t, \bc_I, \bc_T)}_{\substack{\text{image and} \\ \text{text conditioned}}} - \underbrace{e_\theta(\bz_t, \bc_I, \phi_T)}_{\substack{\text{image conditioned}}})}_{\substack{\text{move towards conditioned model} \\ \text{and away from $\bc_I$ conditioned model}}}
    \end{split}
\end{equation}

\begin{table}[!t]
    \centering
    \begin{tabularx}{\columnwidth}{LRRR}
        \toprule
         & \multicolumn{1}{c}{$\times 100$} & \multicolumn{1}{c}{$\times 250$}& \multicolumn{1}{c}{$\times 300$} \\
         \cmidrule(lr){2-2} \cmidrule(lr){3-3} \cmidrule(lr){4-4}
         Model & \lpips & \lpips & \lpips \\
         \midrule
          LRD~\cite{Zhang_2023_ICCV} & \cellcolor{tabsecond}0.2822 &  \cellcolor{tabfirst}0.3013 & \cellcolor{tabsecond}0.3394 \\
        LSID~\cite{Chen_2018_CVPR} &  \cellcolor{tabthird}0.3982 &  \cellcolor{tabthird}0.4342 &  \cellcolor{tabthird}0.4648 \\
        Ours             &  \cellcolor{tabfirst}0.2793 & \cellcolor{tabsecond}0.3032 &  \cellcolor{tabfirst}0.3184\\
        &&\\
         \bottomrule
    \end{tabularx}
    \caption{Quantitative results on the Sony subset of the SID dataset~\cite{Chen_2018_CVPR} in terms of \lpips~\cite{Zhang_2018_CVPR} for the best open-spurce models in the split. The results are conducted on different amplification ratios ($\times 100$, $\times 250$, $\times 300$).}
    \label{tab:res_sid_lpips}
\end{table}
\begin{table}[!t]
    \centering
    \begin{tabularx}{\columnwidth}{LRR}
        \toprule
         & \multicolumn{1}{c}{$\times 100$} & \multicolumn{1}{c}{$\times 200$} \\
         \cmidrule(lr){2-2} \cmidrule(lr){3-3}
         Model & \lpips & \lpips \\
         \midrule
        PMN~\cite{10.1145/3503161.3548186}  & \cellcolor{tabsecond}0.3197 &  - \\
        LRD~\cite{Zhang_2023_ICCV}  &  - & \cellcolor{tabsecond}0.3184 \\
        Ours &  \cellcolor{tabfirst}0.3034 &  \cellcolor{tabfirst}0.3101\\
        &&\\
         \bottomrule
    \end{tabularx}
    \caption{Quantitative results on the SonyA7S2 subset of the ELD dataset~\cite{Wei_2020_CVPR} in terms of \lpips~\cite{Zhang_2018_CVPR} for the best open-spurce models in the split. The results are conducted on different amplification ratios ($\times 100$, $\times 200$).}
    \label{tab:res_eld_lpips}
\end{table}
Our approach is based on fine-tuning standard latent diffusion models like Stable Diffusion~\cite{rombach2022high} which are often trained for the task of text-to-image generation. To support our suite of tasks we add additional channels to the first layer of the underlying UNet, concatenating the latent codes for variably noisy images $\bz_t$ and latent codes for the image conditioning $\mathcal{E}(\bc_I)$~\cite{10.1007/978-3-031-19790-1_26}. We use the same text encoder for the text conditioning vectors $\bc_T$ as in these text-to-image generation models which were trained with text conditioning contrary to the popular approach of replacing the text encoder with an independent image encoder~\cite{justin}. To do so, our approach also relies on generating a set of text prompts for this task and uses these text prompts in addition to the RAW and long-exposure processed image pairs. These text prompts could also potentially be used to guide the model for a variety of other downstream high-level tasks as well.

\begin{figure}[!b]
    \centering
    \begin{tabular}{cc}
        Scaled RAW images & Long-exposure images \\
        \midrule
        \includegraphics[width=0.49\columnwidth]{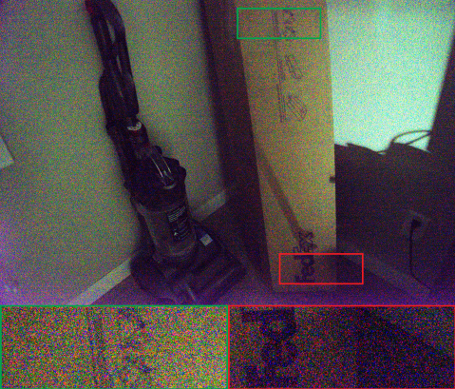} & \includegraphics[width=0.49\columnwidth]{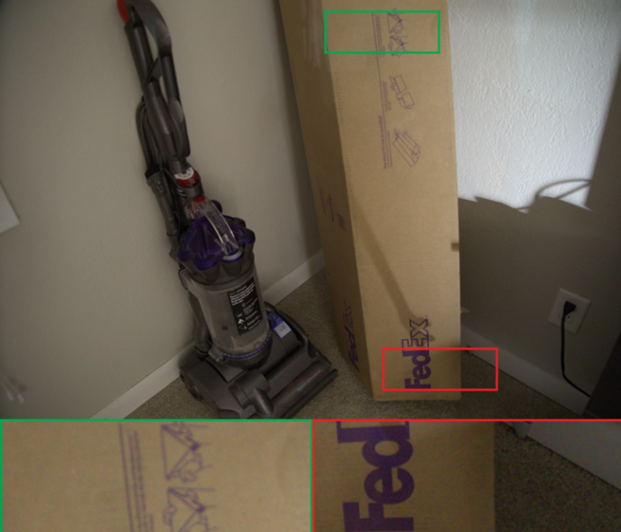}\\
        \includegraphics[width=0.49\columnwidth]{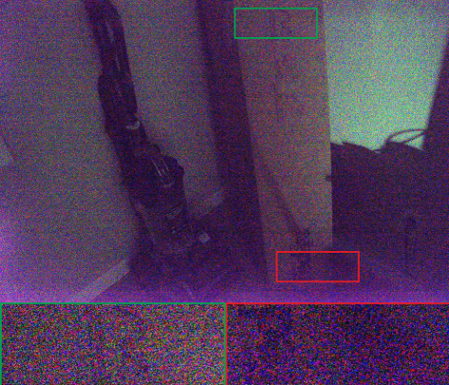} & \includegraphics[width=0.49\columnwidth]{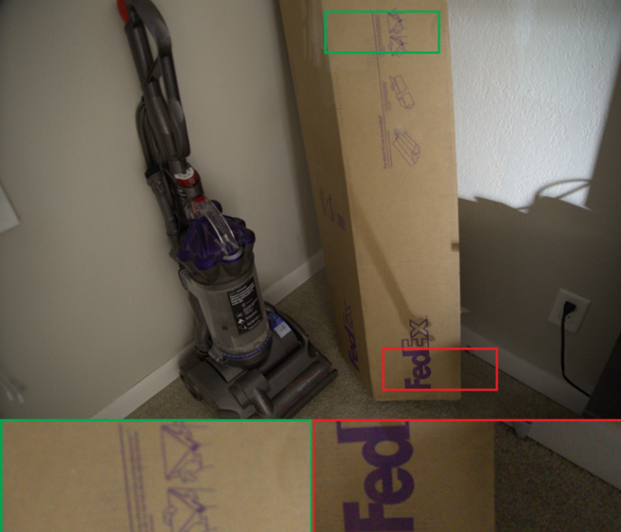}
    \end{tabular}
    \caption{The scaled images and long-exposure ground-truth images for the same scene across three different low-light factors $\times100$ (first row), $\times250$ (second row), and $\times300$ (third row) captured on Sony $\alpha$7SII with an ISO of $10000$.}
    \label{fig:ll-factors}
\end{figure}

\subsection{Training}
\label{sec:training}

\setlength{\tabcolsep}{1.5pt}
\setlength{\fboxrule}{.1pt}
\renewcommand{\arraystretch}{1}
\begin{figure*}[!ht]
    \centering
    \begin{tabular}{cccccc}
         Scaled Images & BM3D~\cite{4271520} & LSID~\cite{Chen_2018_CVPR} & PMN~\cite{10.1145/3503161.3548186} & Ours (\acronym) & Ground Truth\\
         \midrule
         \includegraphics[width=0.16\textwidth]{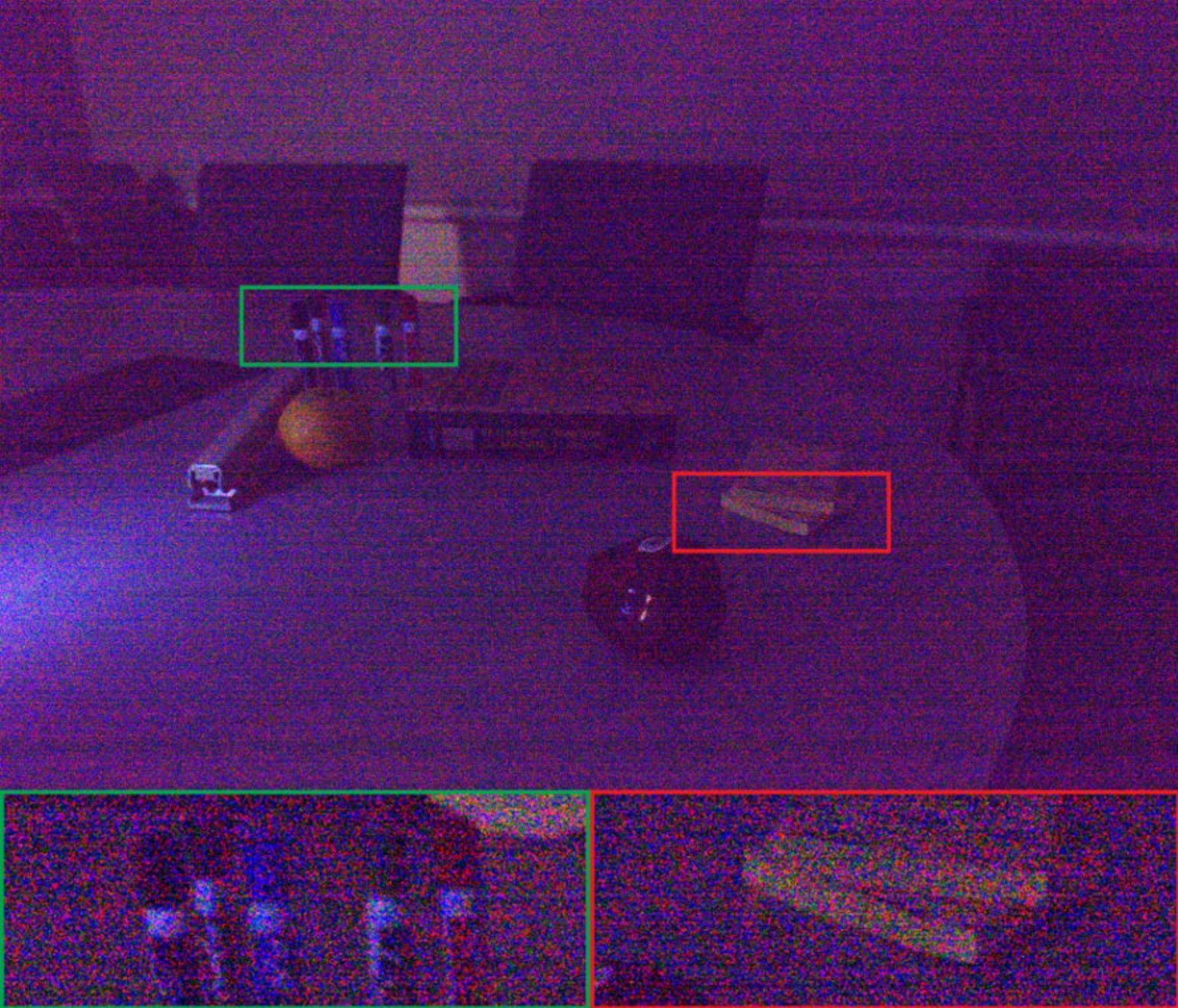} &\includegraphics[width=0.16\textwidth]{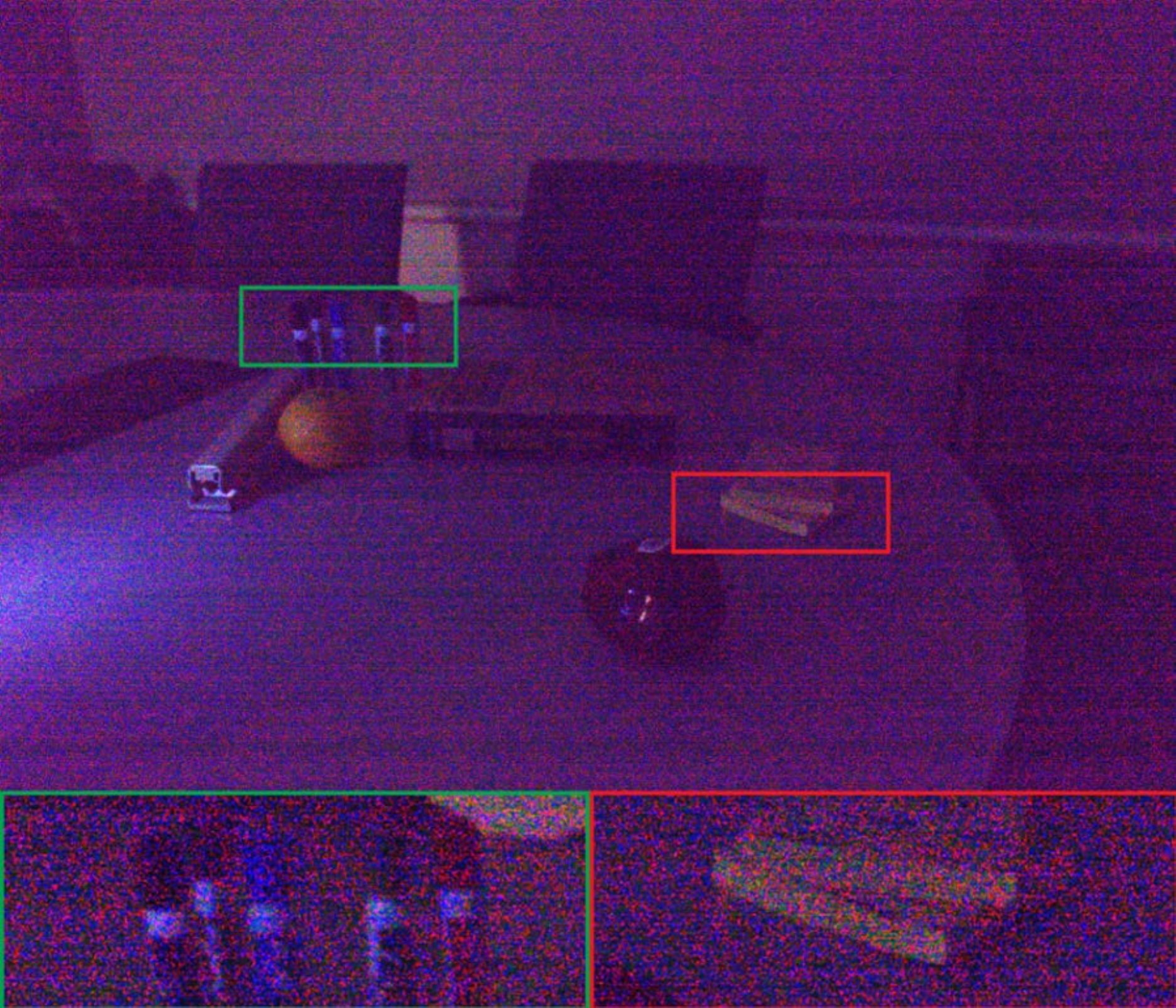}
         &\includegraphics[width=0.16\textwidth]{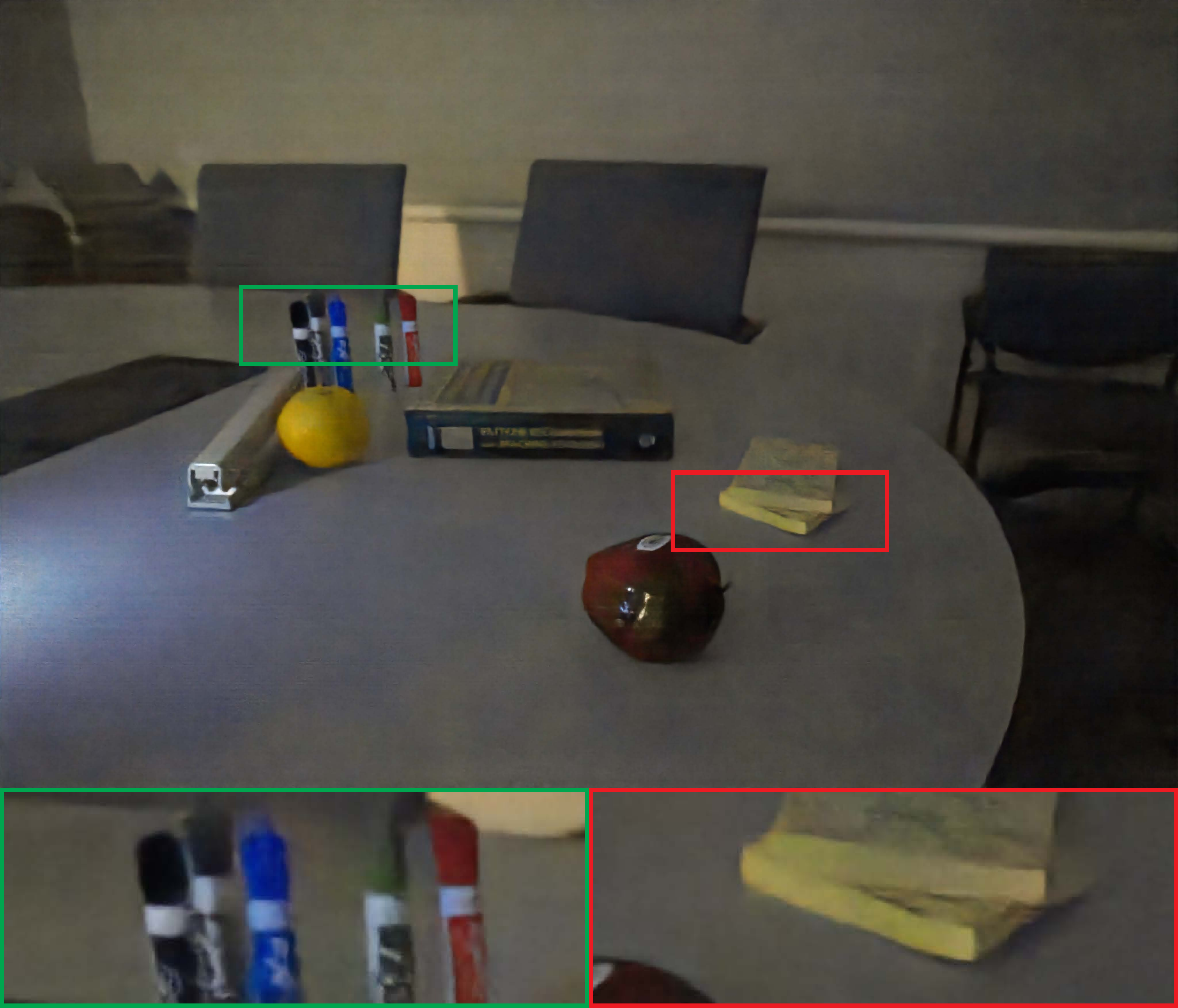} &\includegraphics[width=0.16\textwidth]{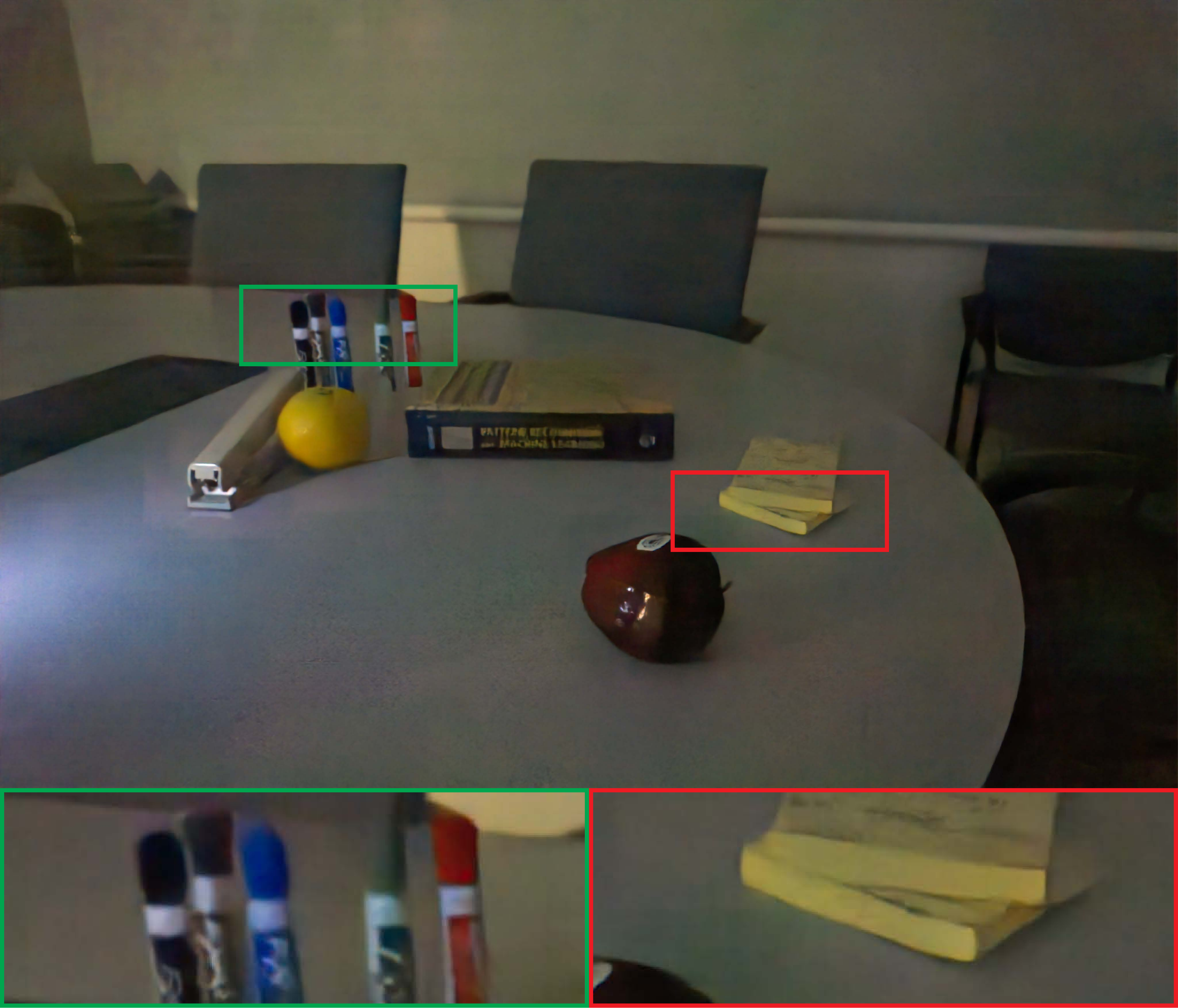} &\includegraphics[width=0.16\textwidth]{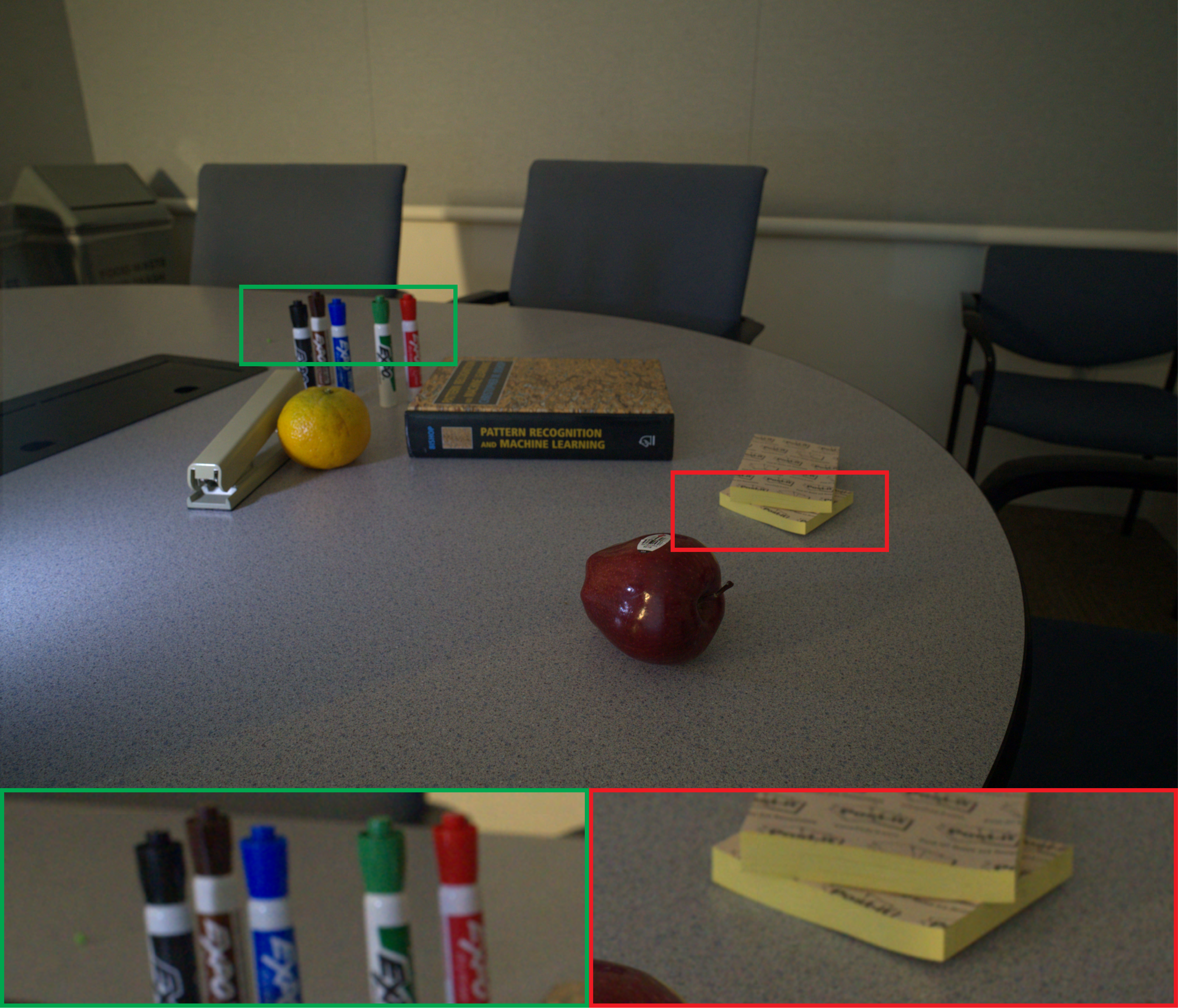} &\includegraphics[width=0.16\textwidth]{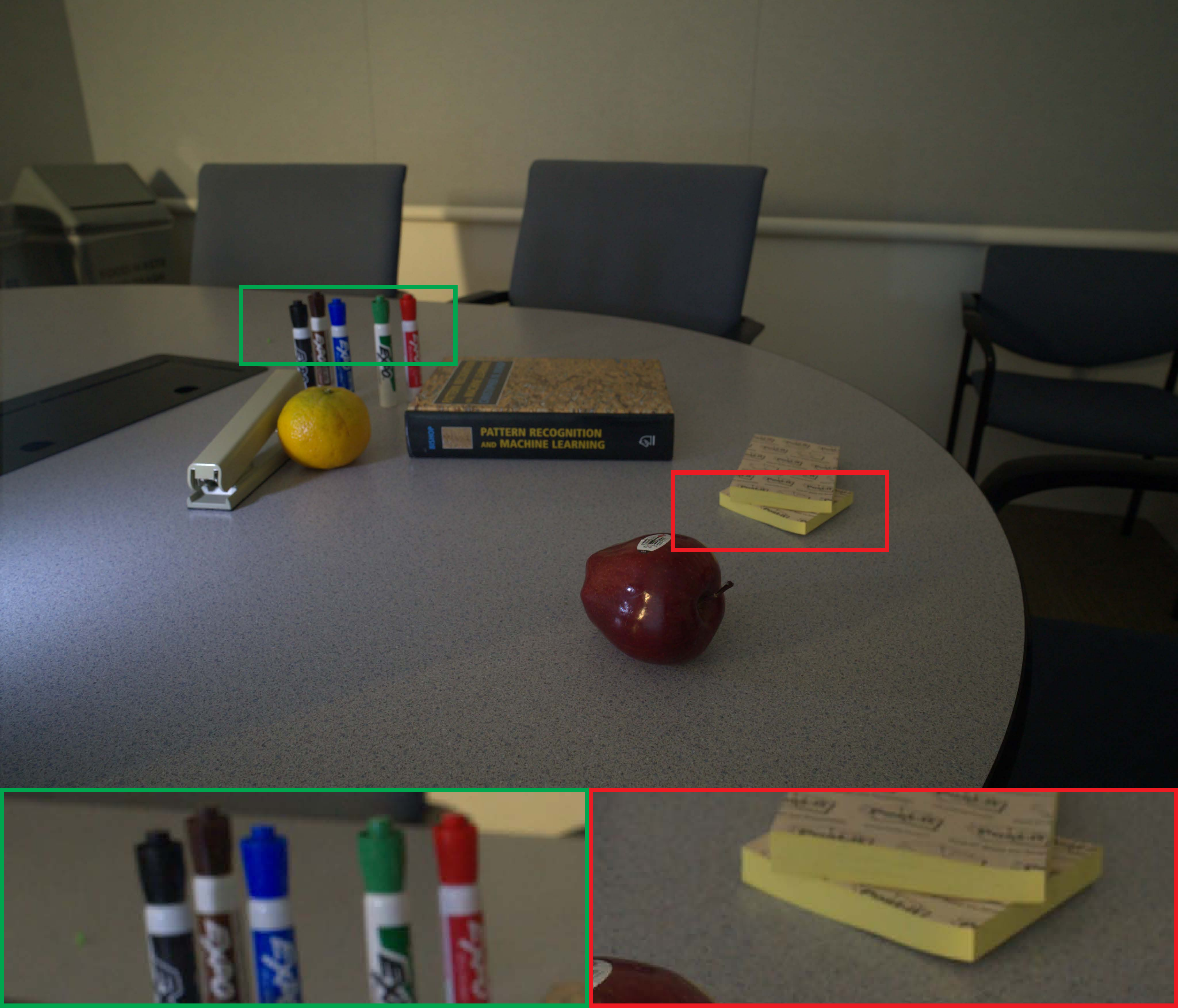} \\
         
         \includegraphics[width=0.16\textwidth]{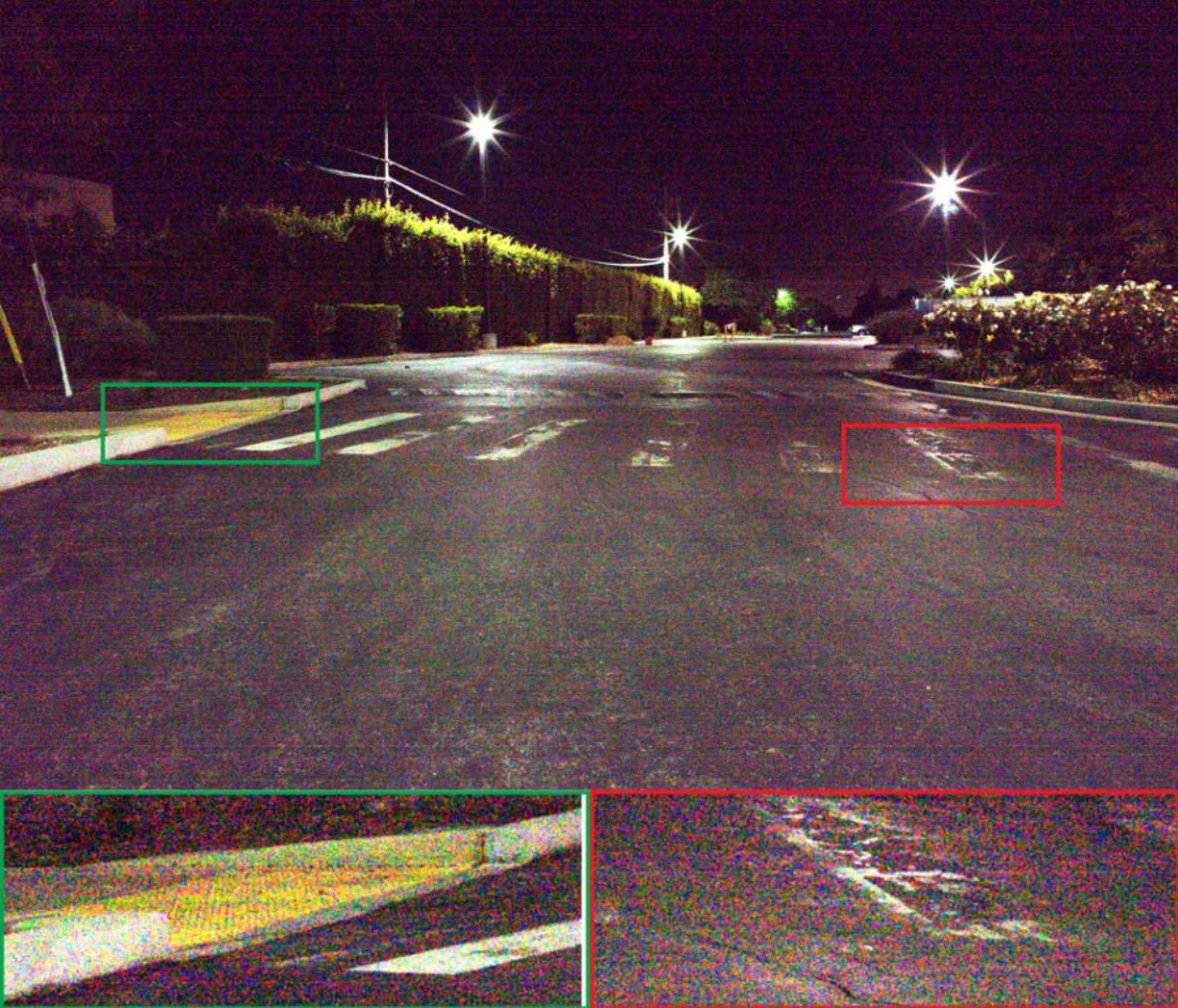} &\includegraphics[width=0.16\textwidth]{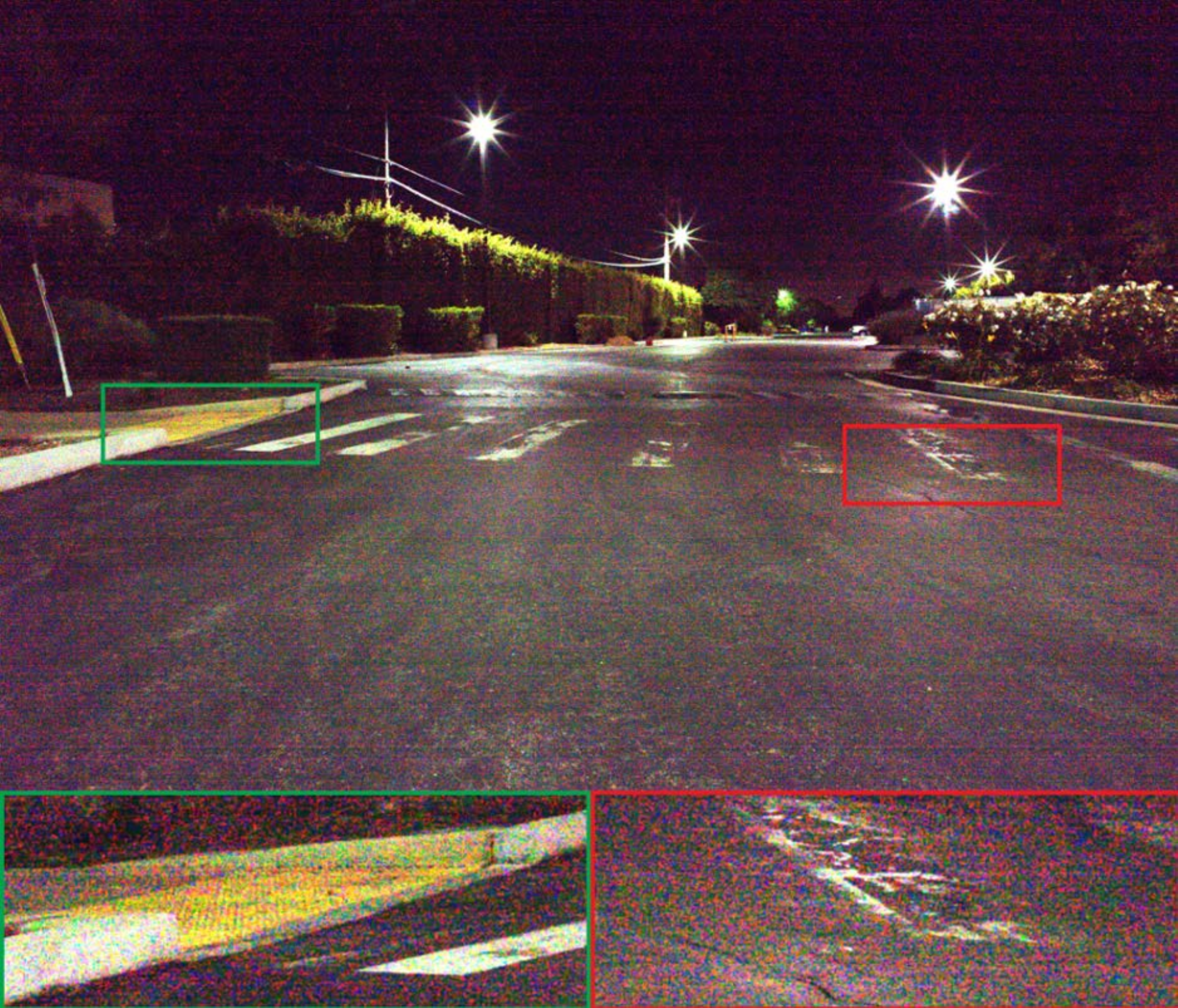}
         &\includegraphics[width=0.16\textwidth]{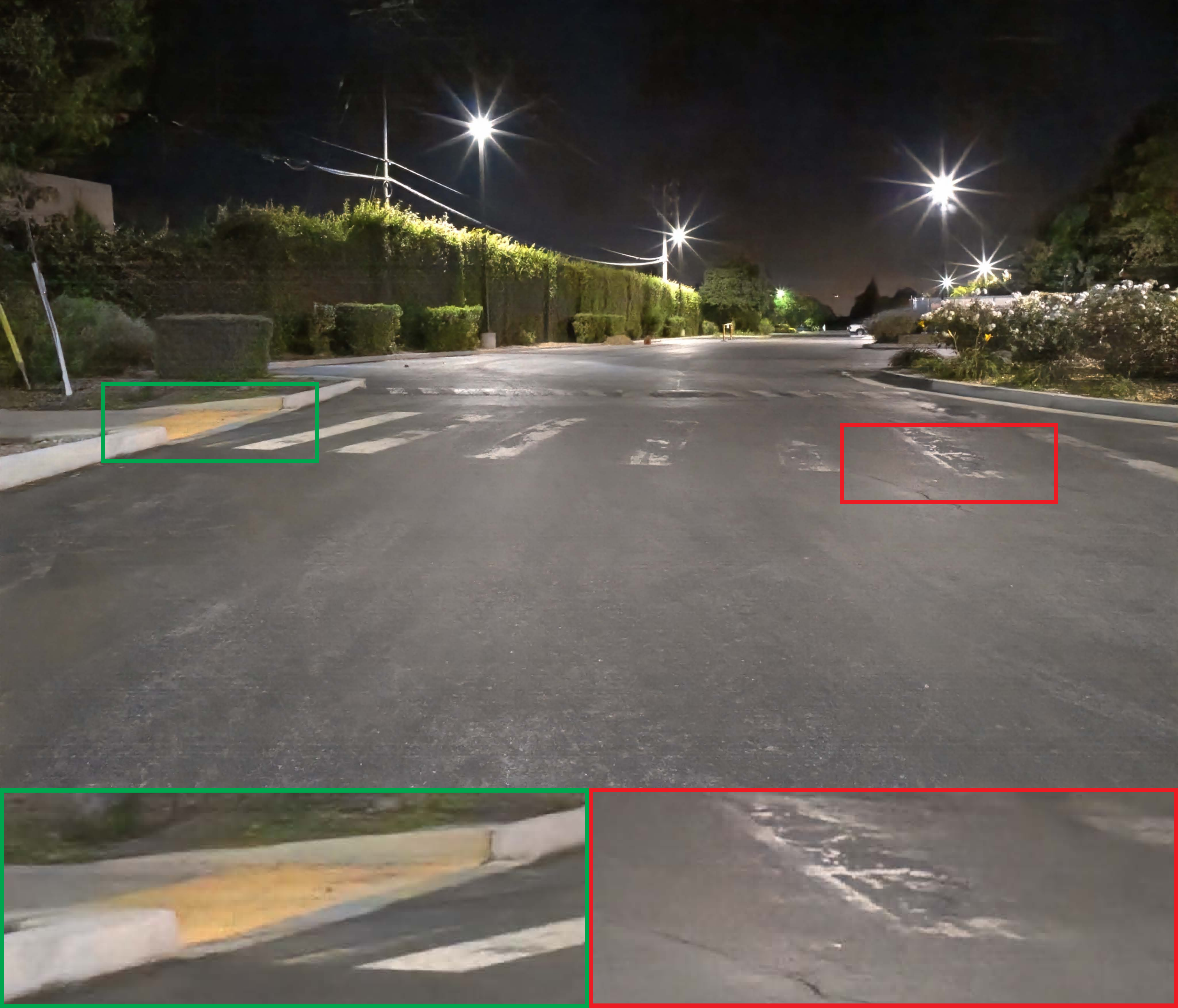} &\includegraphics[width=0.16\textwidth]{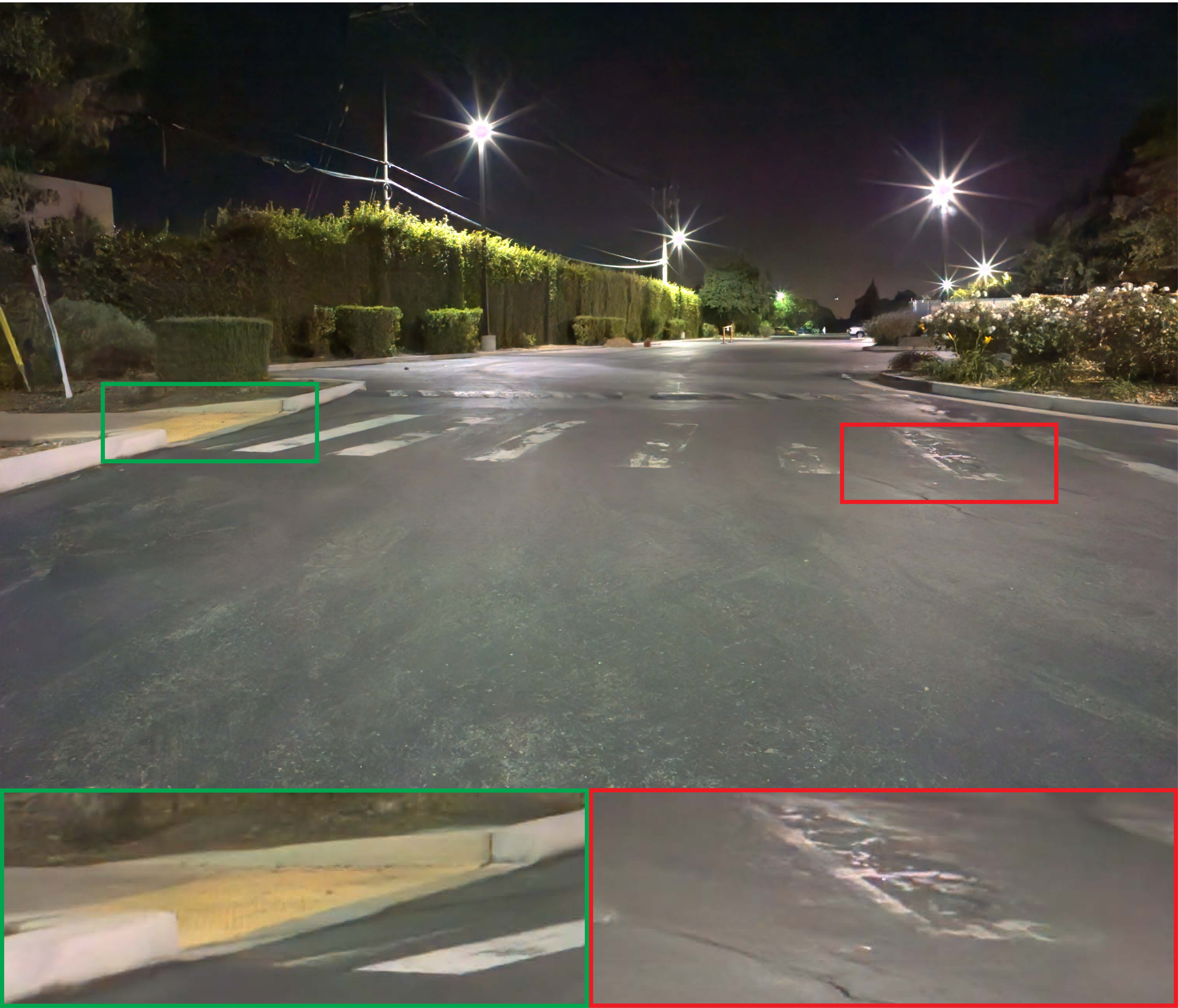} &\includegraphics[width=0.16\textwidth]{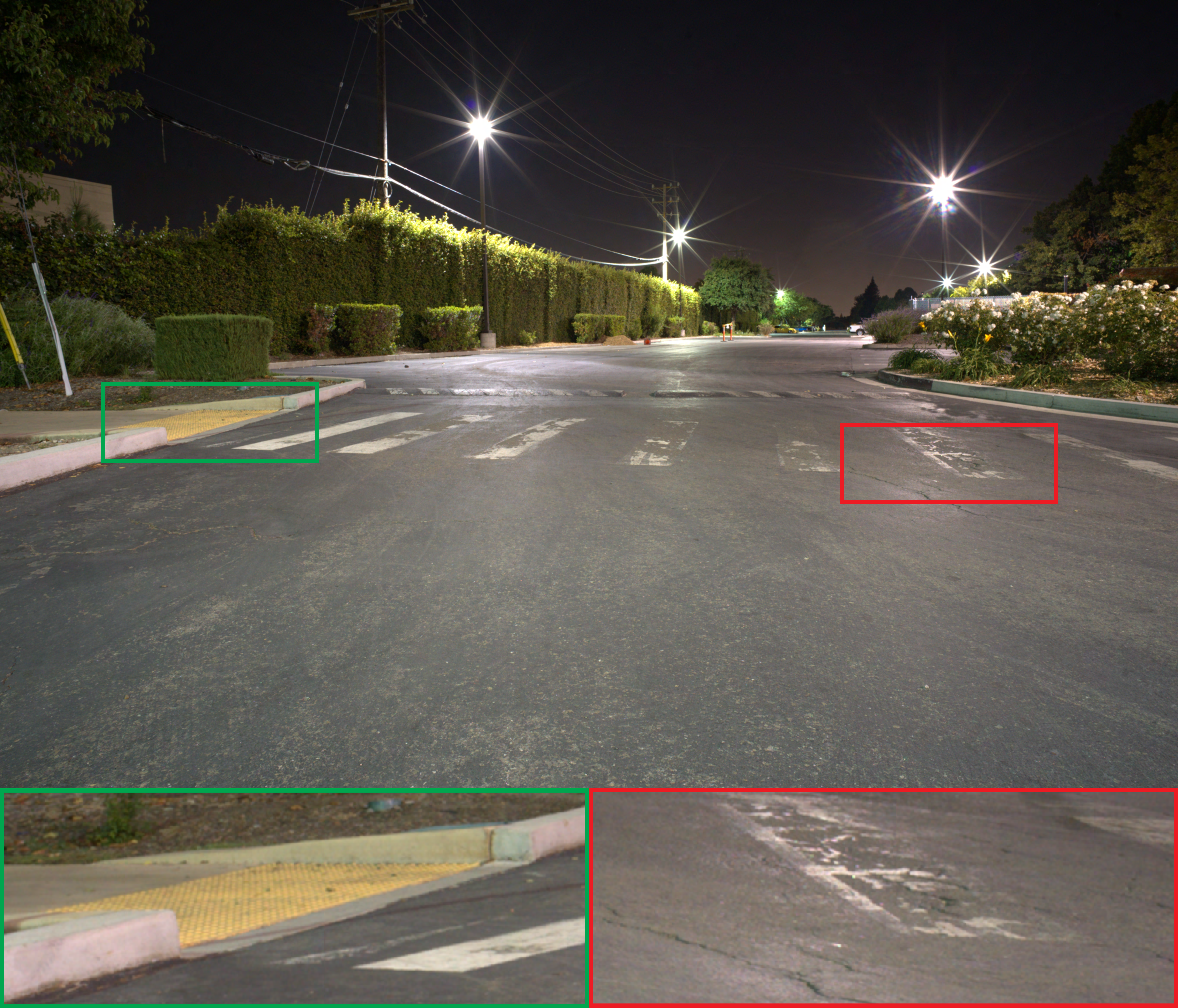} &\includegraphics[width=0.16\textwidth]{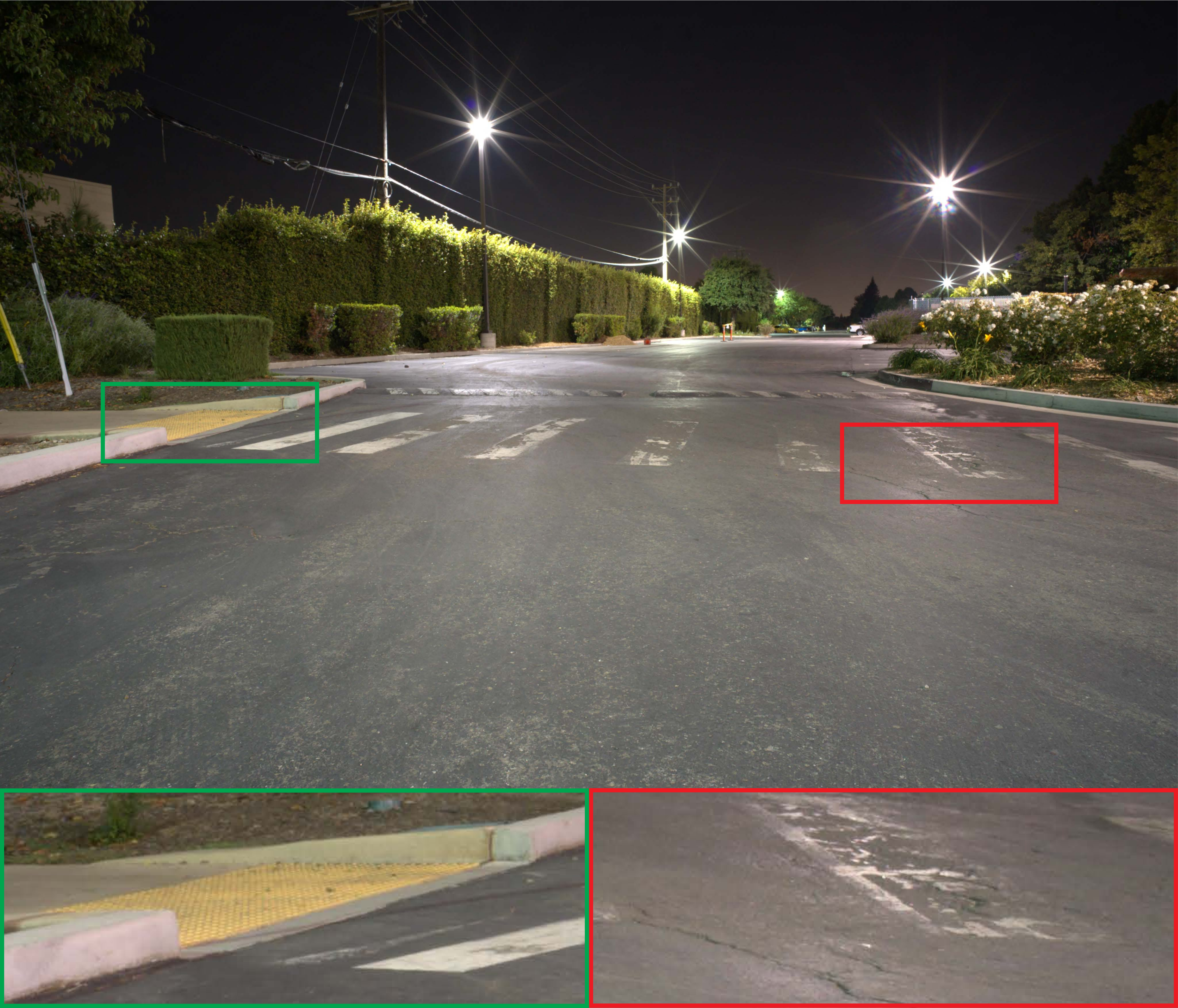} \\
         
         \includegraphics[width=0.16\textwidth]{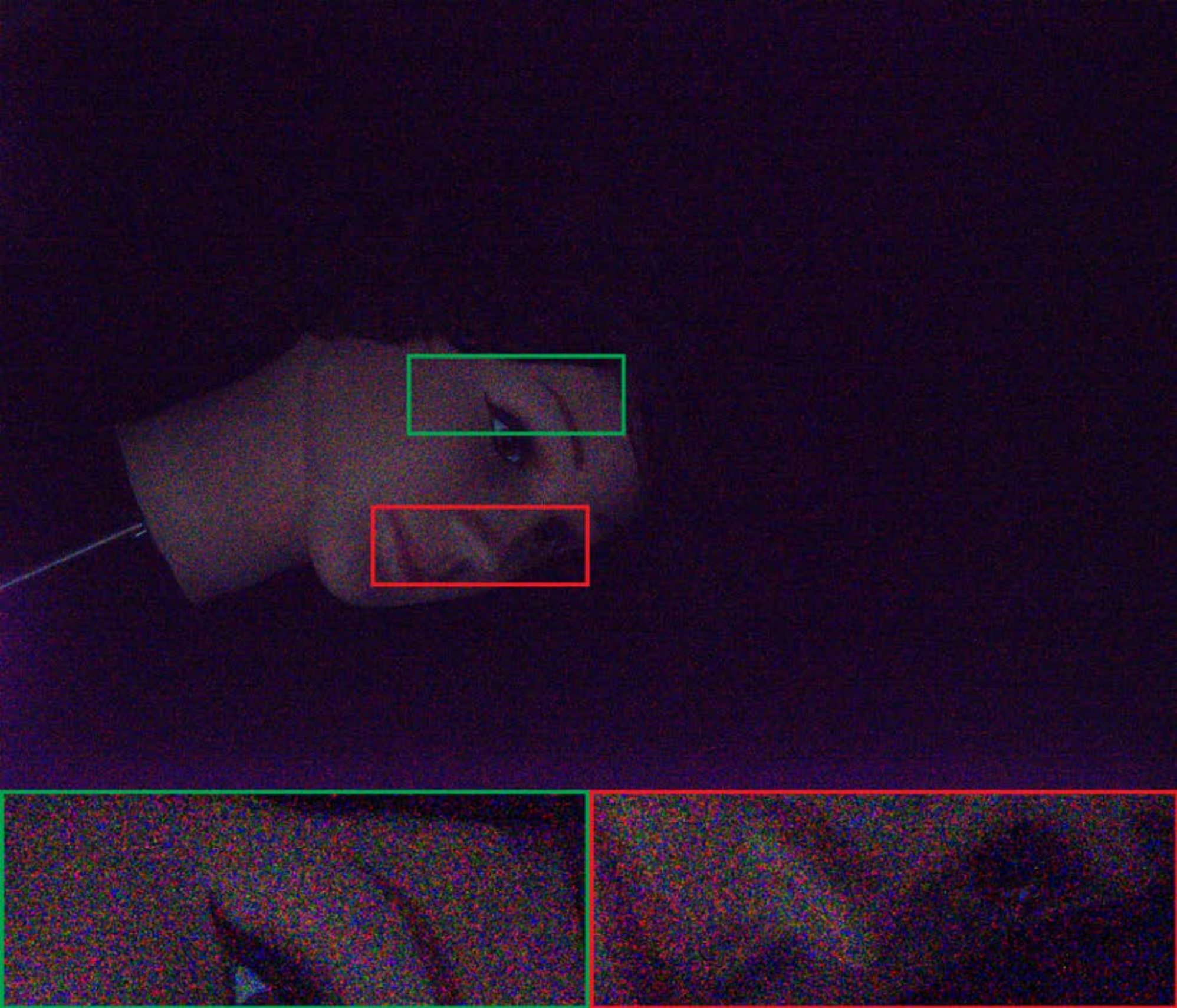} &\includegraphics[width=0.16\textwidth]{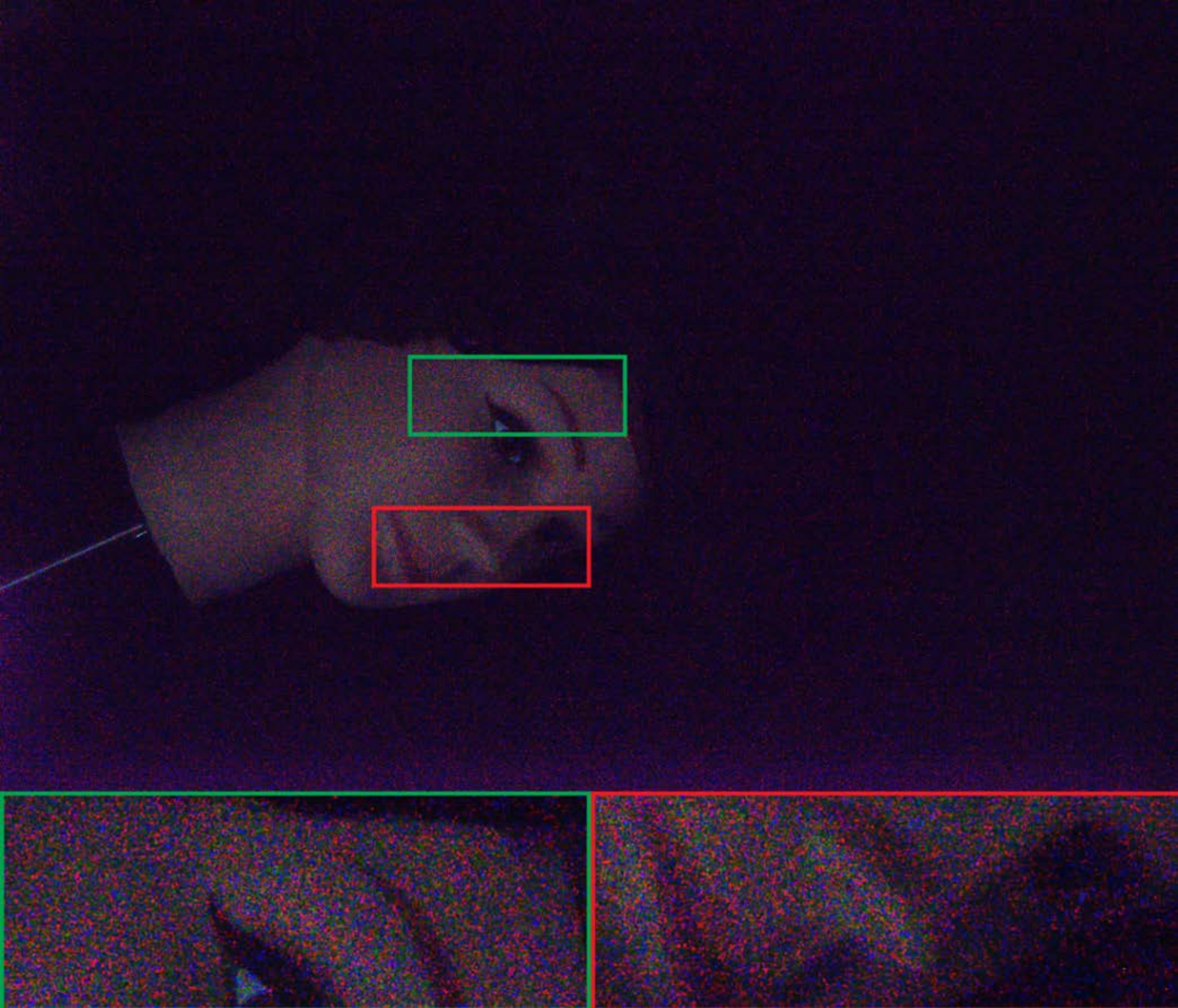}
         &\includegraphics[width=0.16\textwidth]{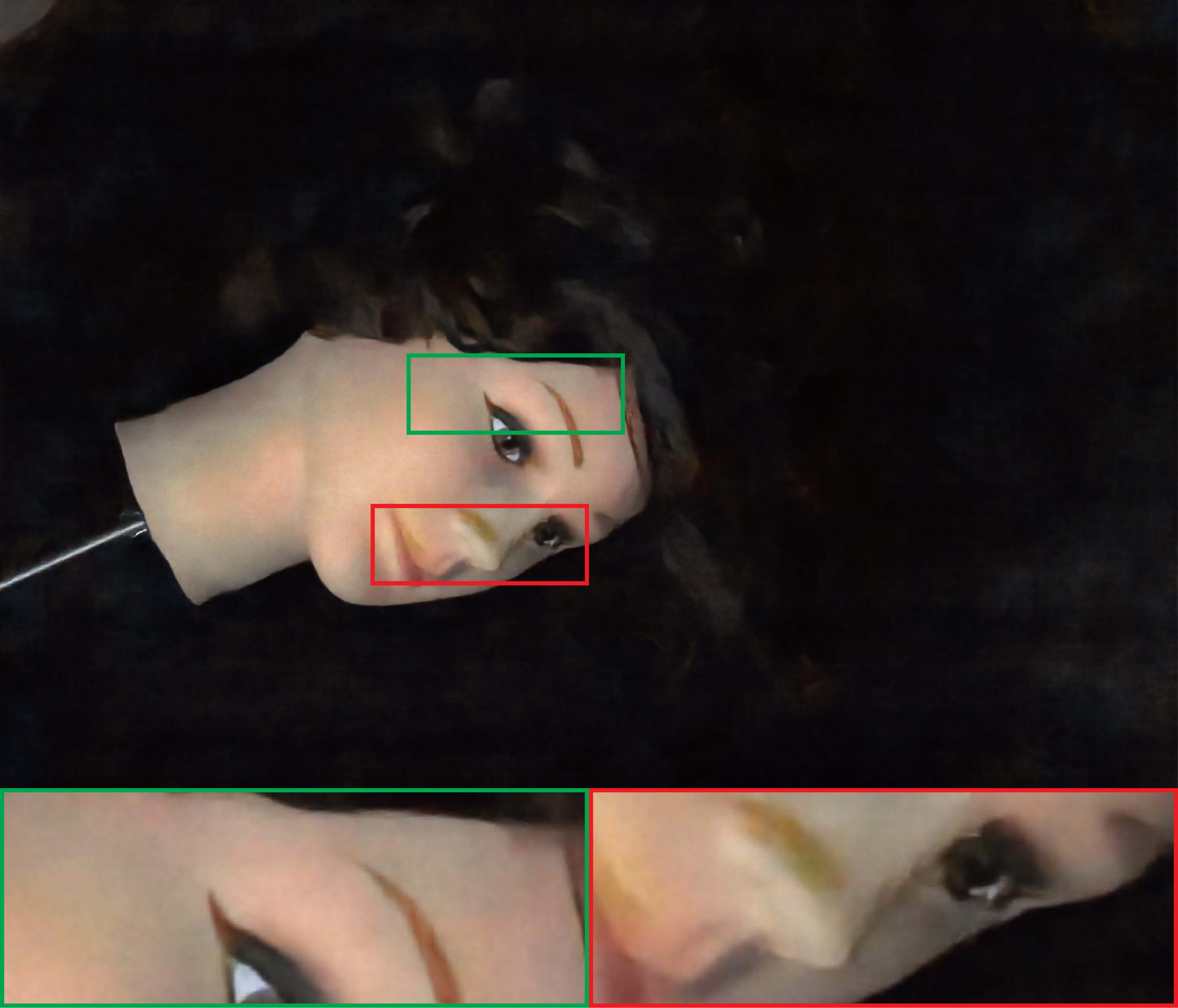} &\includegraphics[width=0.16\textwidth]{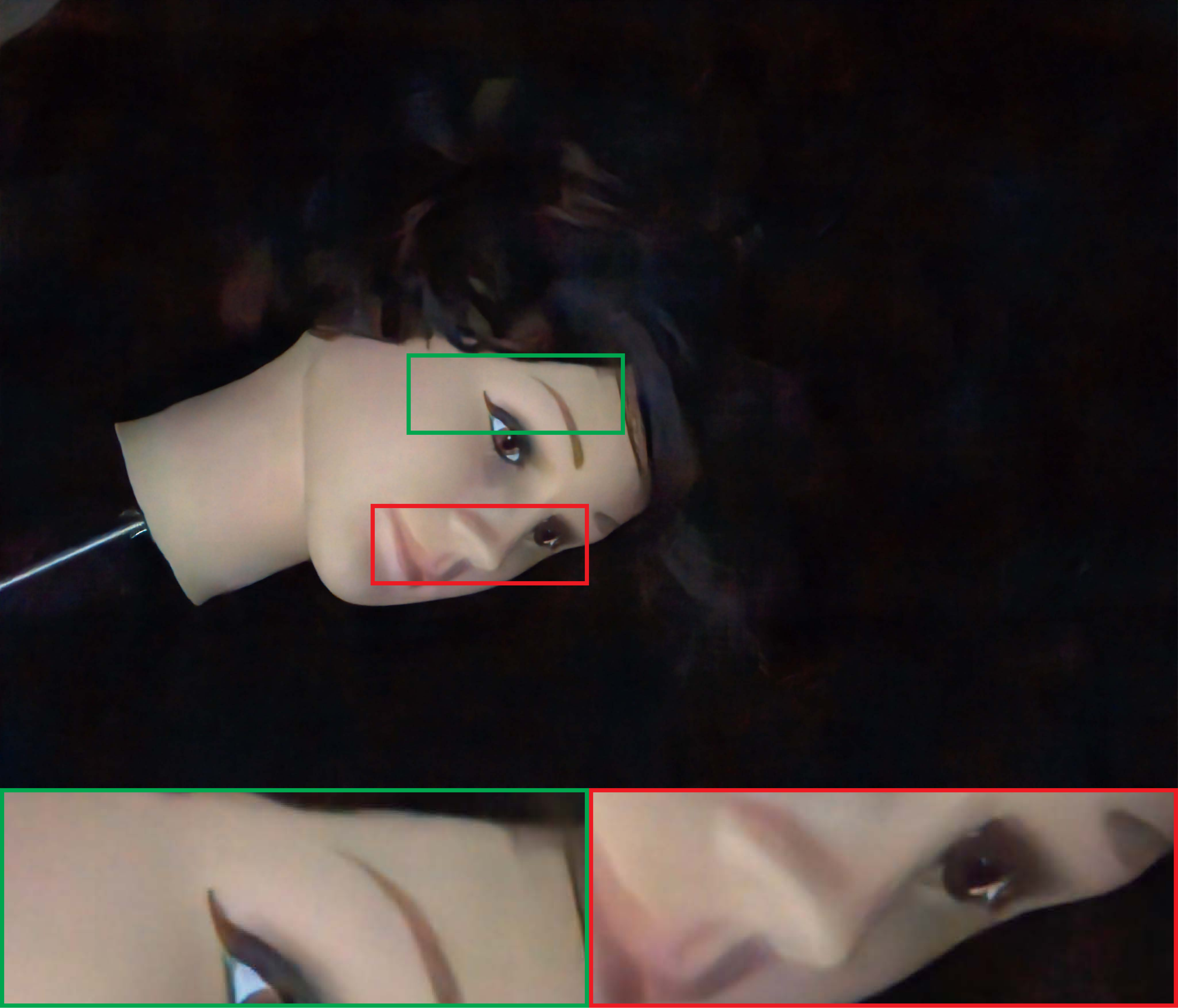} &\includegraphics[width=0.16\textwidth]{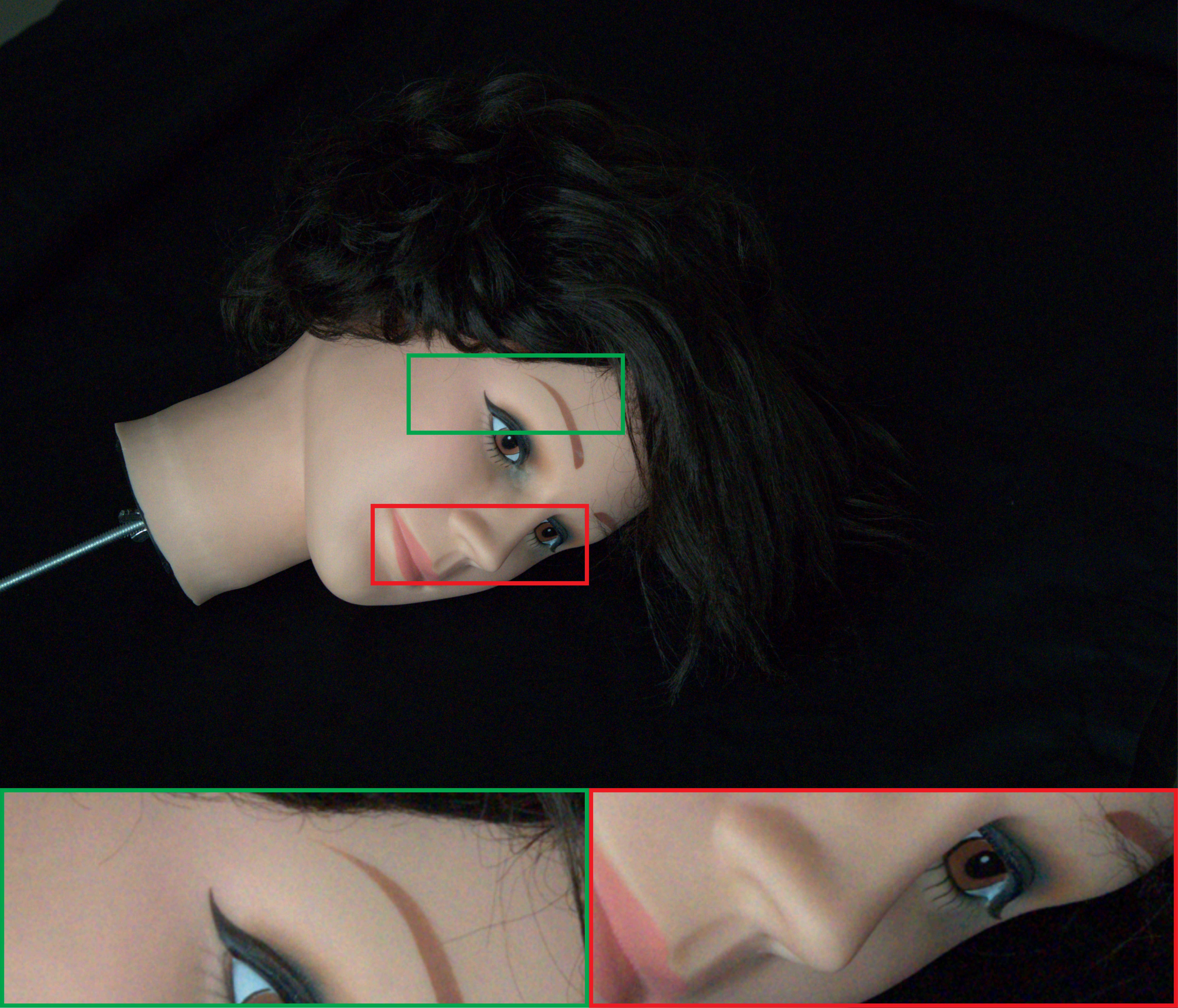} &\includegraphics[width=0.16\textwidth]{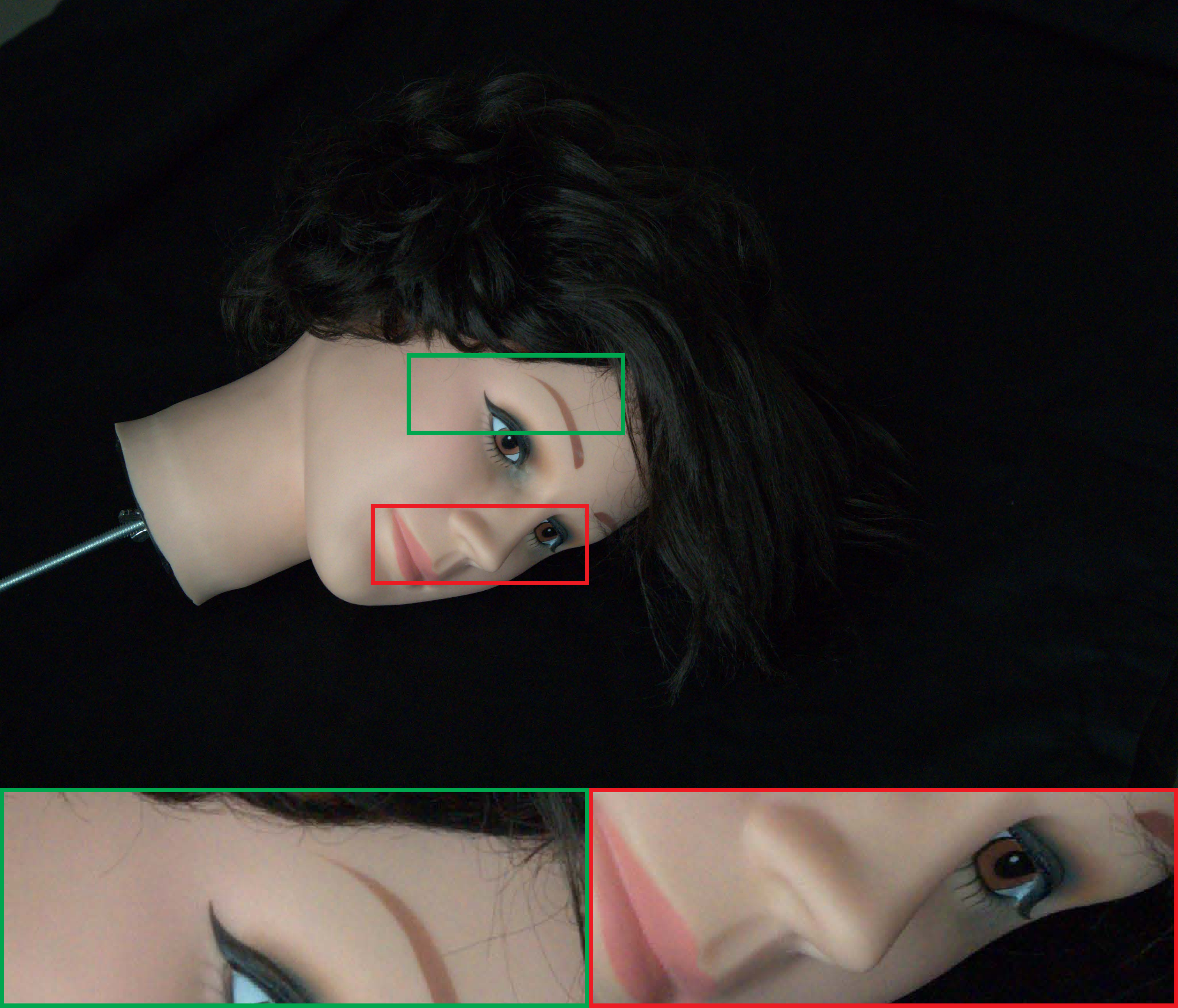}
    \end{tabular}
    \caption{We show results on learning the entire image processing pipeline for extremely low-light enhancement and we compare our approach qualitatively to the best-performing open-source model. These images are taken with Sony $\alpha$7SII and the illuminance at the camera is $<$ 0.1 lux. We show both indoor and outdoor scenes. (\textbf{Best viewed in color and with zoom.})}
    \label{fig:results_comparisions}
\end{figure*}

We fine-tune our models from Stable Diffusion v1.5~\cite{rombach2022high, justin} and fine-tune these models on the See in the Dark (SID)~\cite{Chen_2018_CVPR} and Extreme Low-light Denoising (ELD)~\cite{Wei_2020_CVPR} dataset. We fine-tune these models independently for each camera type. For both of these datasets, while training we use the short-exposure Bayer RAW images which are then packed into four channels (R-Gr-B-Gb) from the dataset, and the corresponding processed long-exposure images (using \texttt{rawpy}~\cite{rawpy}) in the sRGB space. Following~\cite{Chen_2018_CVPR, Wei_2020_CVPR} we set the amplification ratio to be the exposure difference between the input and reference images (we use $\times100$, $\times250$, $\times300$ for the SID dataset and $\times100$, $\times200$ for the ELD dataset) for both training and testing. We train the model on $512^2$ patches which are made as suggested in~\cite{NEURIPS2022_9b013332}.

\section{Experiments}
\label{sec:experiments}

\setlength\tabcolsep{1pt}
\begin{table}[htbp]
    \centering
    \begin{adjustbox}{width=\columnwidth,center}
    \begin{tabularx}{\columnwidth}{LRRRR}
        \toprule
         & \multicolumn{2}{c}{$\times 100$} & \multicolumn{2}{c}{$\times 200$} \\
         \cmidrule(lr){2-3} \cmidrule(lr){4-5}
         Model & \psnr & \ssim & \psnr & \ssim \\
         \midrule
        BM3D~\cite{4271520}               &                      37.69 &                      0.803 &                      34.06 &                      0.696 \\
        N2N~\cite{pmlr-v80-lehtinen18a}               &                      41.63 &                      0.856 &                      37.98 &                      0.775 \\
        P+G~\cite{4623175, Wei_2020_CVPR}                &                      41.76 &                      0.930 &                      39.33 &                      0.872 \\
        NoiseFlow~\cite{Abdelhamed_2019_ICCV}          &                      41.05 &                      0.925 &                      39.23 &                      0.889 \\
        Exposure Diffusion~\cite{Wang_2023_ICCV} &                      43.29 &                      0.929 &                      40.39 &                      0.873 \\
        ELLE~\cite{9428259}              &                      43.11 &                      0.940 &                      40.30 &                      0.884 \\
        Starlight~\cite{Monakhova_2022_CVPR}          &                      43.80 &                      0.936 &                      40.86 &                      0.884 \\
        ELD~\cite{Wei_2020_CVPR}                &                      45.45 &                      0.975 &                      43.43 &                      0.954 \\
       LRD~\cite{Zhang_2023_ICCV}                 &                      44.95 &  \cellcolor{tabthird}0.979 &                      43.32 &                      0.966 \\
        LSID~\cite{Chen_2018_CVPR}              &                      44.47 &                      0.968 &                      41.97 &                      0.928 \\
        PMN~\cite{10.1145/3503161.3548186}                &  \cellcolor{tabthird}46.50 & \cellcolor{tabsecond}0.985 &  \cellcolor{tabthird}44.51 &  \cellcolor{tabthird}0.973 \\
        LLD$^*$~\cite{Cao_2023_CVPR}            & \cellcolor{tabsecond}46.74 &  \cellcolor{tabfirst}0.986 & \cellcolor{tabsecond}44.95 & \cellcolor{tabsecond}0.977 \\
        \midrule
        Ours          &  \cellcolor{tabfirst}47.56 & \cellcolor{tabsecond}0.985 &  \cellcolor{tabfirst}45.12 &  \cellcolor{tabfirst}0.981\\
        &&\\
         \bottomrule
    \end{tabularx}
    \end{adjustbox}
    \caption{Quantitative results on the SonyA7S2 subset of the ELD dataset~\cite{Wei_2020_CVPR} in terms of \psnr \ and \ssim. The results are conducted on different amplification ratios ($\times 100$, $\times 200$).}
    \label{tab:res_eld_psnr_ssim}
\end{table}

\subsection{Implementation Details}

\paragraph{Dataset.} We use noisy-clean pairs from the Sony subset of the SID~\cite{Chen_2018_CVPR} and ELD~\cite{Wei_2020_CVPR} dataset. The images in the Sony subset of the SID dataset which are collected using Sony $\alpha$7SII in 231 static scenes across three low-light factors. There are 1865 image pairs for training, 234 for validation, and 598 for testing. The Sony subset of the ELD dataset consists of 60 low-light image pairs across two low-light factors, which are also captured using the same camera as the SID dataset.

We furthmore, augment all training samples with random crop, random horizontal flip, and use AutoAugment~\cite{cubuk2019autoaugment}. We use the Adam~\cite{kingma2017adam} optimizer, with a learning rate of $5{\rm e}-5$, $500$ warmup steps, $1{\rm e}-2$ weight decay, cosine annealing, and use a batch-size of $64$. We train the model for $300$ epochs. Our code is in PyTorch 2.0~\cite{paszke2019pytorch}. We use a number of open-source packages to develop our training workflows, particularly we use Huggingface Diffusers~\cite{von_Platen_Diffusers_State-of-the-art_diffusion}, Accelerate~\cite{accelerate}, Transformers~\cite{Wolf_Transformers_State-of-the-Art_Natural_2020}, and \texttt{rawpy}~\cite{rawpy}. We utilized mixed-precision training with PyTorch's native AMP (through \texttt{torch.cuda.amp}) for mixed-precision training which allowed us to obtain significant boosts in the overall model training time. Our experiments are conducted on TPUv3-8 and we run test-scripts on Tesla A100.

\subsection{Comparison with state-of-the-art methods}

In order to demonstrate the reliability of our proposed generative camera ISP: \acronym, we conduct experiments on the SID~\cite{Chen_2018_CVPR} and ELD~\cite{Wei_2020_CVPR} datasets for learning the entire image processing pipeline from the RAW data.

\setlength{\tabcolsep}{1.5pt}
\setlength{\fboxrule}{.1pt}
\begin{figure}[!b]
    \centering
    \begin{tabular}{cc}
         \includegraphics[width=0.49\columnwidth]{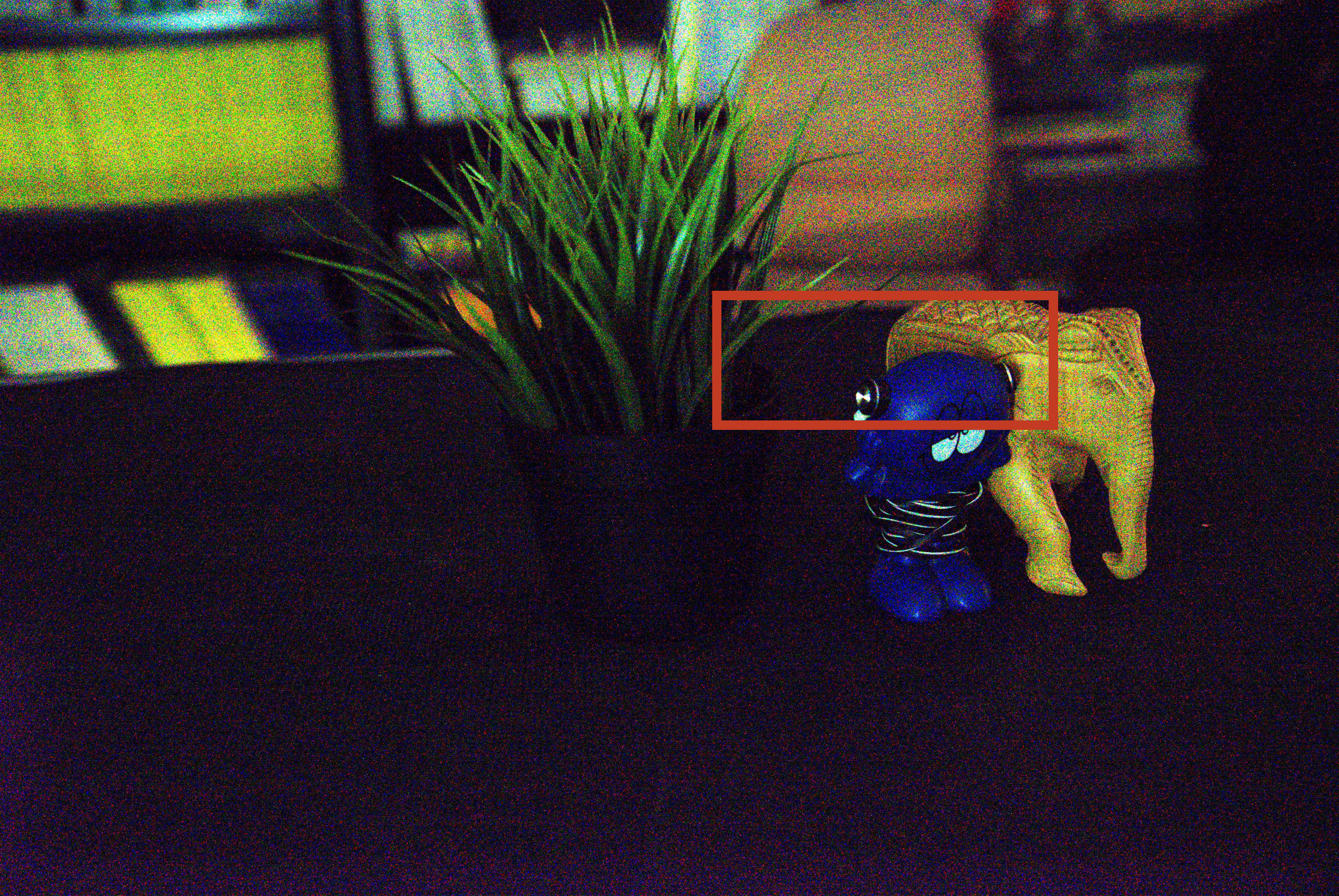} & \includegraphics[width=0.49\columnwidth]{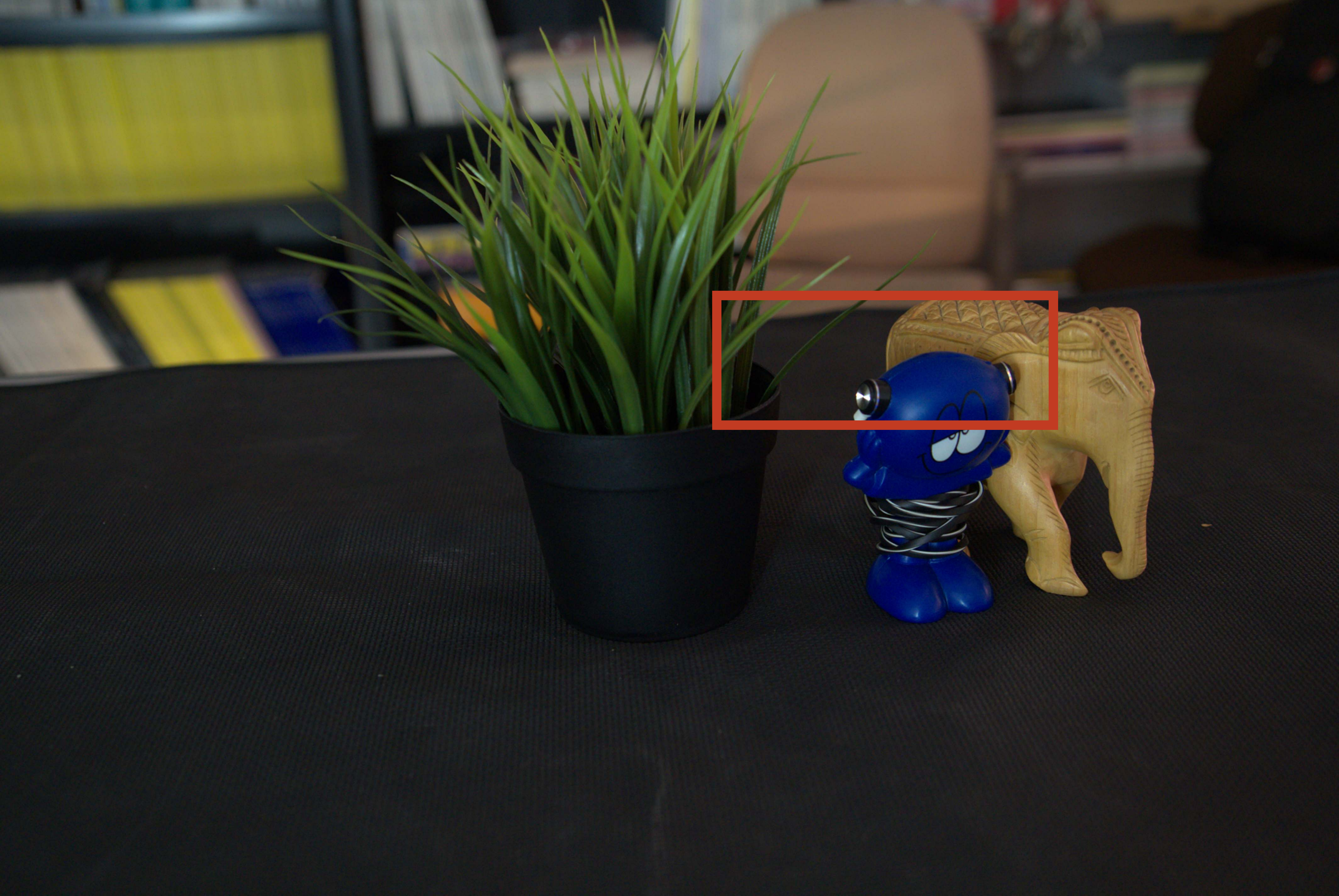} \\
         \includegraphics[width=0.49\columnwidth, cfbox=red 1pt 0pt]{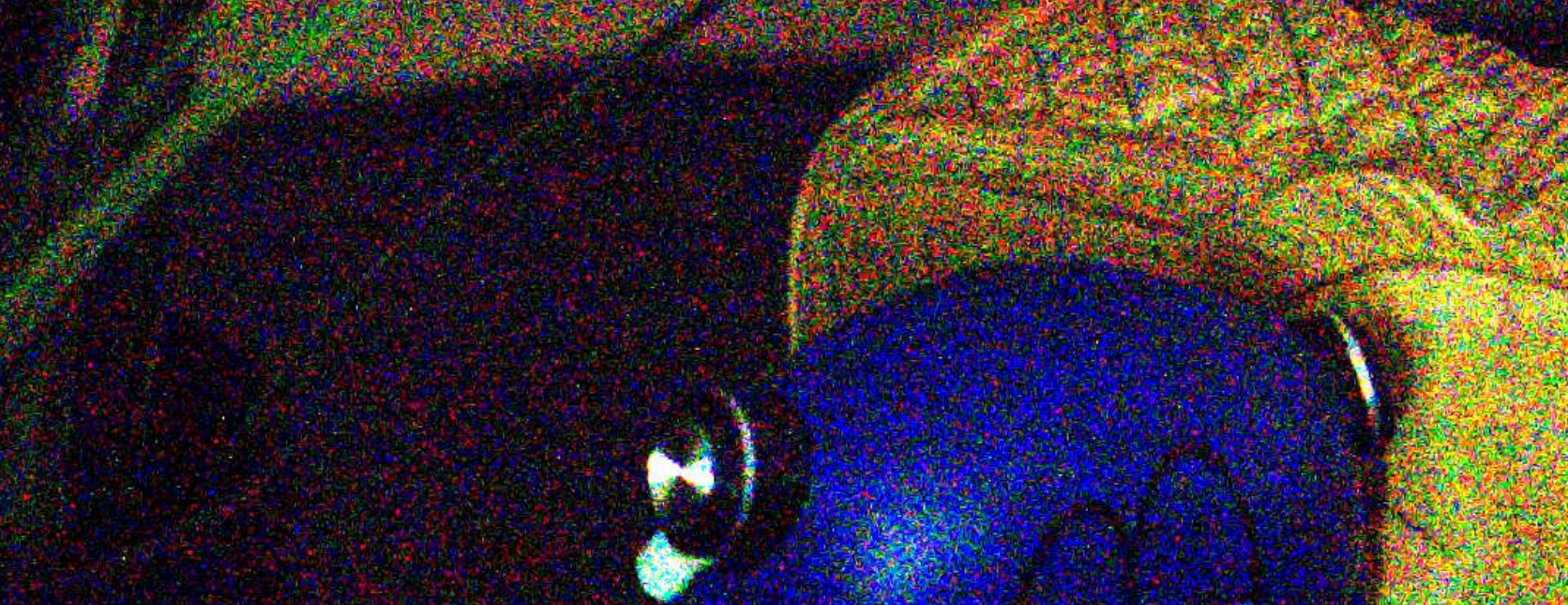} & \includegraphics[width=0.49\columnwidth, cfbox=red 1pt 0pt]{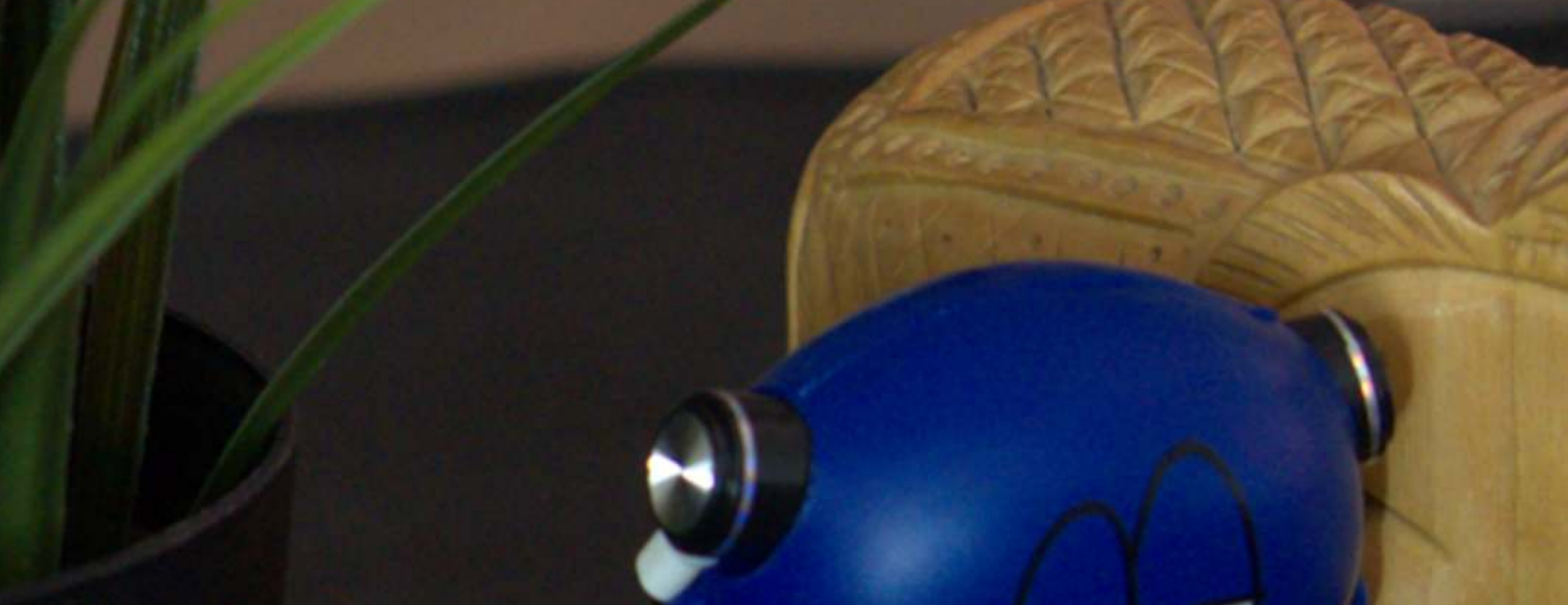} \\
         (a) Traditional Pipeline & (b) Ours (\acronym)
    \end{tabular}
    \caption{We qualitatively compare our model with a traditional pipeline with $\times100$ low-light factor on the SID~\cite{Chen_2018_CVPR} dataset.}
    \label{fig:ours_x100}
\end{figure}

We quantitatively compare our results on the Sony subset of the SID~\cite{Chen_2018_CVPR} dataset with non-learned methods like BM3D~\cite{4271520} and A-BM3D~\cite{5504216}, models based on different noise models~\cite{4623175, Wei_2020_CVPR, pmlr-v80-lehtinen18a}, and neural-net based learning methods~\cite{Abdelhamed_2019_ICCV, Wang_2023_ICCV, Monakhova_2022_CVPR, Chen_2018_CVPR}, which include state-of-the-art models. We summarize these results in Table~\ref{tab:res_sid_psnr_ssim} and Table~\ref{tab:res_sid_lpips} and demonstrate the performance of our approach over different low-light factors. We compare these models based on \psnr \ , \ssim \ , and \lpips~\cite{Zhang_2018_CVPR}. As we can see Table~\ref{tab:res_sid_psnr_ssim} and Table~\ref{tab:res_sid_lpips}, \acronym \ achieves better performance than other state-of-the-art learned methods across most metrics and most low-light factors. We particularly observe that \acronym \ performs significantly better than current state-of-the-art models for $\times300$ low-light factor which is a more challenging task.

\begin{figure}[!b]
    \centering
    \begin{tabular}{cc}
        Scaled Images & Ours (\acronym) \\
        \midrule
        \includegraphics[width=0.46\columnwidth]{lr-figures/noise_level/10226_00_100_scale.png} & \includegraphics[width=0.46\columnwidth]{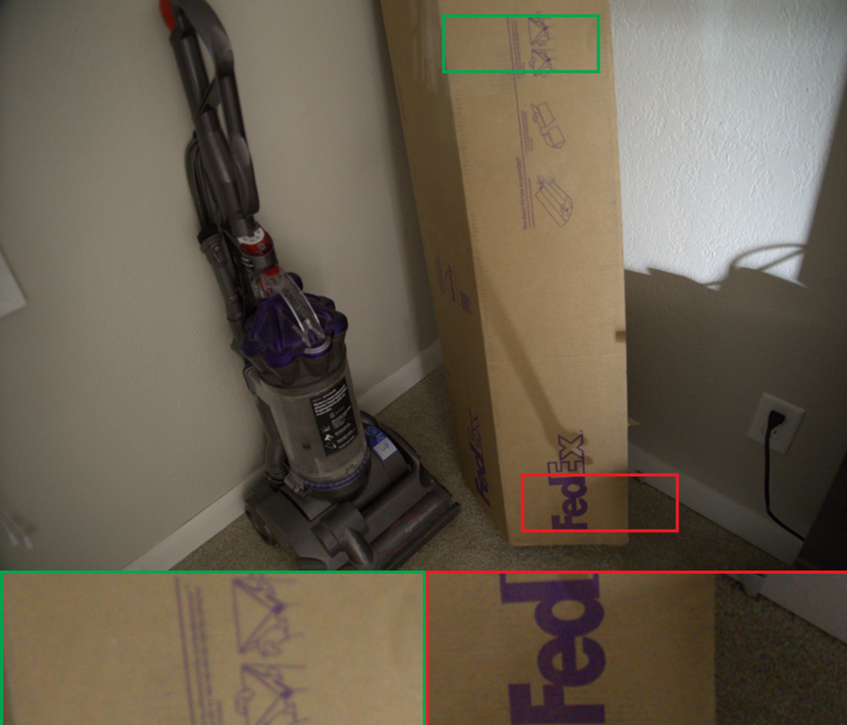}\\
        \includegraphics[width=0.46\columnwidth]{lr-figures/noise_level/10226_00_300_scale.png} & \includegraphics[width=0.46\columnwidth]{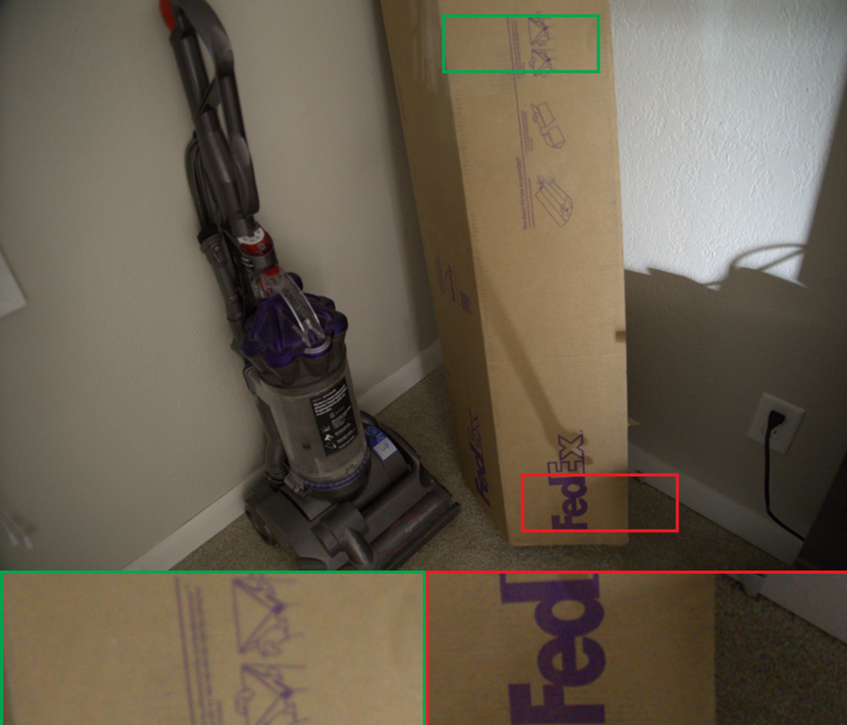}
    \end{tabular}
    \caption{We compare our model on the same scene on different low-light factors for which we observe that the model produces consistent results across different low-light factors, we demonstrate $\times100$ for the first row and $\times300$ for the second row.}
    \label{fig:multiple-factors}
\end{figure}

We also show visual results for our approach and compare it with other approaches in Figure~\ref{fig:results_comparisions}. Though we notice that our model particularly performs very well for $\times250$ and $\times300$, we show the difference between different low-light factors in Figure~\ref{fig:ll-factors}, we show that our model performs well for the $\times100$ task as well which we show qualitatively in Figure~\ref{fig:ours_x100}. We also compare our model on the same scene across low-light factors which we show in Figure~\ref{fig:multiple-factors}.

\setlength{\tabcolsep}{1.5pt}
\setlength{\fboxrule}{.1pt}

We quantitatively compare our results on the Sony subset of the ELD~\cite{Wei_2020_CVPR} dataset with non-learned methods, models based on different noise models~\cite{4623175, Wei_2020_CVPR, pmlr-v80-lehtinen18a}, and neural-net based learning methods~\cite{Abdelhamed_2019_ICCV, Wang_2023_ICCV, Monakhova_2022_CVPR, Chen_2018_CVPR}, which include state-of-the-art models. We summarize these results in Table~\ref{tab:res_eld_psnr_ssim} and Table~\ref{tab:res_eld_lpips} and demonstrate the performance of our approach over different low-light factors. We compare these models based on \psnr \ , \ssim \ , and \lpips~\cite{Zhang_2018_CVPR}. As we can see Table~\ref{tab:res_eld_psnr_ssim} and Table~\ref{tab:res_eld_lpips}, \acronym \ achieves better performance than other state-of-the-art learned methods for $\times200$ low-light factor which is a more challenging task and performs marginally better across the $\times100$ low-light factor in terms of \psnr.
\begin{table*}[tbp]
    \centering
    \begin{tabularx}{\textwidth}{LRRRRRR}
        \toprule
         & \multicolumn{2}{c}{$\times 100$} & \multicolumn{2}{c}{$\times 250$} & \multicolumn{2}{c}{$\times 300$} \\
         \cmidrule(lr){2-3} \cmidrule(lr){4-5} \cmidrule(lr){6-7}
         Model & \psnr & \ssim & \psnr & \ssim & \psnr & \ssim \\
         \midrule
         BM3D~\cite{4271520}               &                      32.92 &                      0.758 &                      29.56 &                      0.686 &                      28.88 &                      0.674 \\
        A-BM3D~\cite{5504216}             &                      33.79 &                      0.743 &                      27.24 &                      0.518 &                      26.52 &                      0.558 \\
        N2N~\cite{pmlr-v80-lehtinen18a}                &                      37.42 &                      0.853 &                      33.48 &                      0.725 &                      32.37 &                      0.686 \\
        P+G~\cite{4623175, Wei_2020_CVPR}                &                      38.31 &                      0.884 &                      34.39 &                      0.765 &                      33.37 &                      0.730 \\
        NoiseFlow~\cite{Abdelhamed_2019_ICCV}          &                      38.89 &                      0.929 &                      35.80 &                      0.867 &                      32.29 &                      0.801 \\
        Exposure Diffusion~\cite{Wang_2023_ICCV} &                      38.89 &                      0.902 &                      36.02 &                      0.832 &                      35.00 &                      0.808 \\
        ELLE~\cite{9428259}               &                      40.09 &                      0.931 &                      36.13 &                      0.863 &                      32.54 &                      0.782 \\
        Starlight~\cite{Monakhova_2022_CVPR}          &                      40.47 &                      0.926 &                      36.25 &                      0.858 &                      32.99 &                      0.780 \\
        ELD~\cite{Wei_2020_CVPR}                &                      41.95 &                      0.953 &                      39.44 &                      0.931 &                      36.36 &                      0.911 \\
        LRD~\cite{Zhang_2023_ICCV}                   &                      41.95 &                      0.956 &                      39.25 &                      0.931 &                      36.03 &                      0.909 \\
        LSID~\cite{Chen_2018_CVPR}               &                      42.06 &                      0.955 &                      39.60 &                      0.938 &                      36.85 &                      0.923 \\
        SFRN~\cite{Zhang_2021_ICCV} & 42.29 & 0.951 & 40.22 & 0.938 & 36.87 & 0.917\\
        PMN~\cite{10.1145/3503161.3548186}                &  \cellcolor{tabthird}43.16 &  \cellcolor{tabthird}0.960 &  \cellcolor{tabthird}40.92 &  \cellcolor{tabthird}0.947 &  \cellcolor{tabthird}37.77 &  \cellcolor{tabthird}0.934 \\
        LLD$^*$~\cite{Cao_2023_CVPR}               & \cellcolor{tabsecond}43.36 & \cellcolor{tabsecond}0.961 & \cellcolor{tabsecond}41.02 & \cellcolor{tabsecond}0.948 & \cellcolor{tabsecond}37.80 & \cellcolor{tabsecond}0.935 \\
        \midrule
        Ours             &  \cellcolor{tabfirst}44.93 &  \cellcolor{tabfirst}0.986 &  \cellcolor{tabfirst}43.46 &  \cellcolor{tabfirst}0.956 &  \cellcolor{tabfirst}39.98 &  \cellcolor{tabfirst}0.951 \\
        &&\\
         \bottomrule
    \end{tabularx}
    \caption{Quantitative results on the Sony subset of the SID dataset~\cite{Chen_2018_CVPR} in terms of \psnr \ and \ssim. The results are conducted on different amplification ratios ($\times 100$, $\times 250$, $\times 300$).}
    \label{tab:res_sid_psnr_ssim}
\end{table*}
\section{Conclusion}
\label{sec:conclusion}

Our method creates a new generative camera ISP for extreme low-light image enhancement that learns the entire image processing pipeline as a diffusion model. Our approach showcases how we can make use of ample image priors in pre-trained text-to-image latent diffusion models for learning tasks in the RAW space. We not only demonstrate the effectiveness of working directly on RAW sensor data for such downstream tasks rather than operating on images processed through some traditional image-processing pipeline but also show the effectiveness of diffusion models to learn in the RAW sensor data space. Parallel to how diffusion models tackle these problems today: learning on processed images, we hope that with this work, we foster further research on using diffusion models and other generative models for tackling the image processing problem and other problems in the RAW space.

\paragraph{Limitations.} Although our approach, \acronym \ generates plausible results, there are several limitations of this work and avenues for future work. An important aspect of performing general low-light camera processing is being able to handle the changing exposure of an image. Secondly, our approach currently does not learn exposure settings well for general scenes or cameras. Our model in its current form is also only able to handle the ISP pipeline for images, an interesting avenue this work could be expanded to would be to explore how we could use similar approaches for video ISP tasks. Finally, we notice, that our approach heavily relies on the use of a proper augmentation regime which could make it challenging to train a model that works for multiple kinds of cameras and learns a universal ISP.
\newpage
\section*{Acknowledgements}

We would like to thank David Lindell of the University of Toronto for multiple insightful discussions. The authors would like to thank Google TPU Research Cloud (TRC) program \footnote{https://sites.research.google/trc/about/} for providing free access to TPUs.

{\small
\bibliographystyle{ieee_fullname}
\bibliography{references}
}

\end{document}